\documentclass[aps,prd,preprint,groupedaddress]{revtex4}
\usepackage{epsfig}
\begin{document}
\def\be{\begin{equation}} 
\def\ee{\end{equation}} 
\def\bea{\begin{eqnarray}} 
\def\eea{\end{eqnarray}} 
\newcommand{\nablaslash}{\nabla \hspace{-0.65em}/}
\title{Low-lying meson spectrum of large $N_C$ strongly coupled lattice QCD}



\author{G.Grignani}
\affiliation{Dipartimento di Fisica and Sezione I.N.F.N. 
Universit\'a di Perugia,
Via A. Pascoli,06123 Perugia, Italy}

\author{D.Marmottini}
\affiliation{Dipartimento di Fisica and Sezione I.N.F.N. Universit\'a di
Perugia, Via A. Pascoli,06123 Perugia, Italy}
\author{P.Sodano}
\affiliation{Dipartimento di Fisica and Sezione I.N.F.N. 
Universit\'a di Perugia,
Via A. Pascoli,06123 Perugia, Italy}

\date{\today}

\begin{abstract}
We compute the low energy mass spectrum  of 
lattice QCD in the large $N_C$ limit.  Expanding around a
gauge-invariant ground state, which spontaneously breaks the discrete chiral 
symmetry, we derive an improved strong-coupling
expansion and evaluate, for any value of $N_C$, the masses of the low-lying
states in the unflavored meson spectrum. We then take the 't Hooft limit
by rescaling $g^2 N_C\to g^2$;
the 't Hooft limit is smooth and no arbitrary parameters are needed. 
We find, already at the fourth order of the strong coupling 
perturbation theory, a very good agreement between the results of our lattice
computation and the known continuum values. 
\end{abstract}


\maketitle
\section{Introduction}

Since the seminal work of 't Hooft~\cite{'tHooft:1973jz,'tHooft:1974hx} 
the large $N_C$ limit (with $g^2 N_C$ 
fixed) has played an increasingly important role in studying gauge 
theories in the
continuum and on the lattice~\cite{Teper:2002kh} and, more recently,
 the duality between gauge and string theories
~\cite{Aharony:1999ti,Kruczenski:2003be}.  
Our aim is to use the 't Hooft limit to investigate some features of the 
meson spectrum of strongly coupled lattice QCD.
Our results exhibit a good agreement with the
observed meson masses, thus implying that in this context also
the 't Hooft limit offers a very accurate method to analyze the QCD spectrum.

A quite successful approach to studying gauge theories with
confining spectra is the strong-coupling
expansion~\cite{Wilczek:1999id}.
In the strong coupling limit, confinement is explicit, the
confining string is a stable object~\cite{Wilson:1974sk}, and some
other qualitative features of the spectrum are easily
obtained. The formulation of the strong coupling expansion requires a
gauge invariant ultraviolet cutoff, which is most conveniently
implemented using lattice regularization. One of the most
difficult problems of the strong-coupling approach to lattice gauge
theory remains its extrapolation to the continuum limit, which usually
occurs at weak coupling. 
In spite of this difficulty, there are strong-coupling
computations that claim a high degree of success.  A useful test of 
strong-coupling expansions in lattice gauge theories has been the study of
lower dimensional models
whose solution in the continuum is known even in the strong coupling
regime; for the Schwinger models~\cite{Berruto:1997jv} and the 
two-dimensional 't Hooft model~\cite{Berruto:1999ir,Berruto:2002gn}, 
the meson spectrum 
and the chiral condensate have 
been evaluated using the strong-coupling approximation 
and a remarkable agreement with the known exact results was found.
Furthermore, it is by now well known~\cite{Smit:1980nf} that relevant 
features of strongly coupled lattice gauge theories have analogues in quantum 
spin systems: in many instances selecting the appropriate 
ground state for building the strong-coupling expansion of a gauge model
turns out to be equivalent to 
finding the ground state of a generalized quantum 
antiferromagnet~\cite{Langmann:hd}. This equivalence turns out to provide 
pertinent nonperturbative information about the structure of the gauge 
invariant states and, in many instances, greatly simplifies the computation of
the chiral condensate.

Many choices of strong-coupling theory produce identical continuum
physics. As it is well known, any irrelevant operator may be added
to the lattice Hamiltonian with an arbitrary parameter. Even if
such ad hoc terms do not modify the continuum limit their coefficients 
might appear in physical quantities such as the hadron 
masses~\cite{Banks:1976ia}.
Our study evidences that the 't Hooft limit does not 
require any arbitrary parameter and leads naturally to unambiguous
and well defined results.

We shall develop a method enabling us to extend to a general number 
of colors the
celebrated approach to the computation of the meson spectrum 
in strongly coupled lattice QCD first introduced by Banks et
al. in Ref.~\cite{Banks:1976ia}.
We shall use the Hamiltonian
approach to lattice gauge theory using staggered 
fermions~\cite{Grignani:2003ix}. 
The staggered formalism is known to yield good results
in the strong-coupling evaluation of the hadron spectrum~\cite{Banks:1976ia};
in contrast, other types of lattice fermions such as
domain-wall or overlap fermions at strong coupling are expected to
suffer both doubling and explicit breaking of 
chiral symmetry~\cite{Brower:1999ak}.
In the staggered fermion formalism, due to fermion doubling, 
the number of continuum fermions
with $N_f$ lattice fermions in $D$ space-time dimensions is $N_f 2^{[D/2]-1}$.
If, as in this paper, one includes only one lattice fermion, 
the continuum limit yields the two lightest quarks 
$(u,d)$. In this paper we shall determine the spectrum 
of the low-lying unflavored mesons.

The computations are performed using $x=1/g^2$ as the expansion
parameter and the extrapolation to the continuum limit is carried
by means of Pad\'e approximants. 
As the ground state we take the one of the pertinent 
antiferromagnetic Ising model. The gauge invariant eigenstates 
of the unperturbed Hamiltonian are used to develop the perturbative expansion.
The 't Hooft limit is then taken: the coupling 
constant is rescaled according to $g^2N_C\to g^2$
and then $N_C$ is sent to infinity.
As shown in~\cite{Witten:1979kh} mesons for large $N_c$ are free,
stable and non-interacting and their masses have smooth limits.
Zweig's rule is exact at large $N_C$, mixing of mesons with 
glue states are suppressed and mesons for large $N_C$ are 
pure $q\bar{q}$ states.
The meson energies computed perturbatively in the strong-coupling expansion
contain terms depending on the number of lattice links. 
In the thermodynamic limit these cancel against
the ground state vacuum energy. Thus the results we shall present are
finite, parameter independent and already in good agreement with the 
known physical values at the fourth order in the perturbative 
strong-coupling expansion.

In Sec. II we review the well known
Hamiltonian formulation of lattice QCD with staggered fermions and
provide a classification of the symmetries of the theory.

In Sec. III we
consider the strong-coupling limit of lattice QCD and determine the
chiral symmetry breaking ground state. We then evaluate its
energy up to the fourth order in the strong-coupling
expansion.

In Sec. IV we construct the operators creating
mesons from the vacuum and then compute their energies up to the
fourth order in the perturbative expansion. We show that these energies
are finite and well defined in the large $N_C$ limit.

Section V is devoted to the extrapolation of the lattice results to the
continuum theory. There we show that the lattice theory, properly
extrapolated to the continuum via Pad\'e approximants,already well reproduces
the known experimental values for the ratios between meson masses
at low orders in the strong-coupling expansion.

Section VI is devoted to some concluding remarks, while the appendices
illustrate some technical aspects needed to clarify the computations
reported in the paper.

\section{Lattice QCD in the Hamiltonian formulation}

In the Hamiltonian formulation of lattice QCD
with staggered fermions~\cite{Susskind:1976jm}
time is a continuous variable and space is discretized on a
$3$-dimensional cubic lattice with $M$ sites, labeled by $\vec{r}=(x,y,z)$;
with $x,y$ and $z$ integers. The conventions and the notation used in
our paper are briefly reviewed in Appendix A. 

The lattice QCD Hamiltonian with one lattice flavor of massless
quarks may be written as the sum of three contributions
\begin{equation}
\label{A}
H=H_e+\tilde{H}_q+H_m,
\end{equation}
where\begin{eqnarray}
\label{H}
&& H_e=\frac{g^2}{2a}\sum_{[\vec{r},\hat{n}]}E^a[\vec{r},\hat{n}]^2\\
\label{I}
&&
\tilde{H}_q=\frac{1}{2a}\sum_{[\vec{r},\hat{n}]}\eta(\hat{n})\Psi_A^{\dagger}(\vec{r}
+\hat{n})U_{AB}[\vec{r},\hat{n}]\Psi_B(\vec{r})+\text{H.c.}\equiv
H_q+H_q^{\dagger}\\
\label{L}
&&
H_m=\frac{1}{2g^2a}\sum_{[\vec{r},\hat{n},\hat{m}]}\left\{Tr(U[\vec{r},
\hat{n}]U[\vec{r}+\hat{n},\hat m]U^{\dagger}[\vec{r}+\hat{m},\hat n]
U^{\dagger}[\vec{r},\hat{m}])+\text{H.c.}\right\}
\end{eqnarray}
are the electric field Hamiltonian, the interaction Hamiltonian between 
quarks and gauge fields and the magnetic Hamiltonian, respectively.
The sums $\sum_{[\vec{r},\hat{n}]}$ are extended to the $N$ lattice 
links, whereas $\sum_{[\vec{r},\hat{n},\hat{m}]}$ is a sum over the
plaquettes.
$\hat{n}=\hat{x},\hat{y},\hat{z}$ is the unit vector in the $\vec n$ direction 
and
\begin{equation}
\label{B}
\eta(\hat{x})=(-1)^z, \qquad \eta(\hat{y})=(-1)^x, \qquad
\eta(\hat{z})=(-1)^y
\end{equation}
\noindent 
are the Dirac $\vec\alpha$ matrices for staggered fermions
~\cite{Susskind:1976jm}.\\
\indent The gauge field
$U[\vec{r},\hat{n}]$ is associated with the link $[\vec{r},\hat{n}]$
and it is a group element in the fundamental representation of
$SU(N_C)$. Two gauge fields occupying the same link are related by
\begin{equation}
\label{D}
U[\vec{r},\hat{n}]=U^{\dagger}[\vec{r}+\hat{n},-\hat{n}].
\end{equation}
\noindent 
The electric field operator $E^a[\vec{r},\hat{n}]$ is
defined on a link and it obeys the algebra
\begin{equation}
\label{E}
\left[E^a[\vec{r},\hat{n}],E^b[\vec{r^{\prime}},\hat{m}]\right]
=if^{abc}E^c[\vec{r},\hat{n}]\delta([\vec{r},\hat{n}]
-[\vec{r^{\prime}},\hat{m}]).
\end{equation}
\noindent 
$E[\vec{r},\hat{n}]=E^a[\vec{r},\hat{n}]T^a$, where
$T^a$, $a=1,...,N_C^2-1$, are the generators of the Lie algebra of
$U(N_C)$. They generate the left-action of the Lie algebra on
$U[\vec{r},\hat{n}]$~\cite{Langmann:hd,Luo:1998dx}
\begin{eqnarray}
\label{F}
&&
\left[E^a[\vec{r},\hat{n}],U[\vec{r^{\prime}},\hat{m}]\right]=
-T^aU[\vec{r},\hat{n}]
\delta([\vec{r},\hat{n}]-[\vec{r^{\prime}},\hat{m}])\\
\label{F1}
&& \left[E^a[\vec{r},\hat{n}],U^{\dagger}[\vec{r^{\prime}},\hat{m}]\right]=
U^{\dagger}[\vec{r},\hat{n}]T^a\delta([\vec{r},\hat{n}]-[\vec{r^{\prime}},
\hat{m}]).
\end{eqnarray}
\noindent The fermion fields $\Psi$ are defined on the lattice sites and 
obey the anticommutation relations
\begin{eqnarray}
\label{G}
&& \{\Psi_A(\vec{r}),\Psi^{\dagger}_B(\vec{r^{\prime}})\}=
\delta_{AB}\delta(\vec{r}-\vec{r^{\prime}})\\
\nonumber
&& \{\Psi_A(\vec{r}),\Psi_B(\vec{r^{\prime}})\}=0
\end{eqnarray}

In addition to gauge invariance,the lattice Hamiltonian (\ref{A}) 
is invariant under the following discrete symmetries 
\begin{enumerate}
\item Lattice translation by even integers,
\begin{equation}
\Psi(x,y,z)\rightarrow\Psi(x+2l,y+2p,z+2q)
\end{equation}
\noindent where $l$, $p$, $q$ are integers.
This symmetry operation amounts to a discrete 
translational invariance on the lattice.
\item Lattice translation by a single link,
\begin{eqnarray}
\label{transl}
&& \Psi(r)\rightarrow\Psi(r+\hat{x})(-1)^y\\
\nonumber
&& \Psi(r)\rightarrow\Psi(r+\hat{y})(-1)^z\\
\nonumber
&& \Psi(r)\rightarrow\Psi(r+\hat{z})(-1)^x.
\end{eqnarray}
In momentum space the last equation can be written as
$$q\to e^{ik_z}\gamma_{5}\tau_{3}q$$
which in the continuum limit, where $k_z$ is infinitesimal, gives
$$q\to \gamma_5\tau_3q;$$
the other two transformations yield
\begin{eqnarray}
\nonumber
&& q\to \gamma_5\tau_2q\\
\nonumber
&& q\to \gamma_5\tau_1q.
\end{eqnarray}
The invariance under translation by a single
link plays then the role of a discrete chiral invariance of the
theory. This symmetry is broken by an explicit mass term in the
Hamiltonian. As we shall show later, the chiral symmetry on the lattice
is also spontaneously broken by the vacuum, and this should
generate a nonvanishing chiral condensate $<\bar\psi\psi>$.
\item Shift along a face diagonal,
\begin{eqnarray}
&& \Psi(r)\rightarrow(-1)^{x+y}\Psi(r+\hat{x}+\hat{z})\\
\nonumber
&& \Psi(r)\rightarrow(-1)^{y+z}\Psi(r+\hat{y}+\hat{x})\\
\nonumber
&& \Psi(r)\rightarrow(-1)^{z+x}\Psi(r+\hat{z}+\hat{y}).
\end{eqnarray}
These transformations correspond to the discrete isospin rotations
\begin{eqnarray}
\nonumber
&& q\to \tau_2 q\\
\nonumber
&& q\to \tau_3 q\\
\nonumber
&& q\to \tau_1 q.
\end{eqnarray}
\item Parity,
\begin{equation}
\Psi(r)\rightarrow\Psi(-r).
\end{equation}
\noindent This is the reflection through the origin.
\item G-parity
\begin{equation}
\Psi(r)\rightarrow\Psi^{\dagger}(-r).
\end{equation}
\noindent This is just complex conjugation.
\end{enumerate}
\noindent

\section{The strong-coupling limit and the vacuum state energy}

In this section we shall describe in some detail the method we propose
for implementing the strong-coupling expansion for 
QCD with a generic number of colors $N_C$, by explicitly 
computing the vacuum state energy up
to the fourth order in the strong coupling expansion.

In the strong-coupling expansion the electric field Hamiltonian $H_e$, 
Eq.(\ref{H}), 
is the unperturbed Hamiltonian while the interaction
Hamiltonian between quarks and gauge fields $H_q$, Eq.(\ref{I}), 
and the magnetic
Hamiltonian $H_m$, Eq.(\ref{L}), are treated as perturbations.
Furthermore, each term in $H$, Eq.(\ref{A}), is gauge invariant since
\begin{equation}
\label{L1}
[\mathcal{G}^a(\vec{r}),H_i]=0~,~~~~i=e,q,m.
\end{equation}
\noindent In Eq.(\ref{L1})
\begin{equation}
\mathcal{G}^a(\vec r) =\sum_{i=-\vec{n}}^{\vec{n}}E^a[\vec r,\hat i]+\Psi_A^{\dagger}(\vec r)T^a_{AB}\Psi_B(\vec r)
\end{equation}
are the generators of static gauge transformations and 
$a=1,...,N_C^2-1$; these generators obey the Lie
algebra
\begin{equation}
[\mathcal{G}^a(\vec{r}),\mathcal{G}^b(\vec{r^{\prime}})]=if^{abc}\mathcal{G}^c(\vec{r})\delta(\vec{r}-\vec{r^{\prime}}).
\end{equation}
\noindent The empty vacuum $|0>$ is a gauge-invariant 
($\mathcal{G}^a(\vec r)|0>=0$) singlet of the electric field
algebra ($E^a[\vec r,\hat i]|0>=0$) which contains no fermions 
($\psi_A(\vec r)|0>=0$). 

Due to gauge invariance, $|0>$ must also be color singlet and should 
be charge neutral, i.e., it should obey the equation 
$\sum_{\vec{r}}\rho(\vec{r})|0>=0,$
with $\rho(\vec{r})$ given by the local fermion number operator
\be
\label{rhoop}
\rho(\vec{r})=\frac{1}{2}[\Psi^{\dagger}(\vec{r}),\Psi(\vec{r})]=
\Psi^{\dagger}(\vec{r})\Psi(\vec{r})-\frac{N_C}{2}.
\ee
\noindent At a given site one may form a color singlet in two ways: one can
leave the site unoccupied or one may put on it $N_C$ fermions with
antisymmetrized singlet wave function. Fermi statistics allow at most
one singlet per site and it is possible to distribute the $M/2$ singlets
arbitrarily ($M$ is the total number of lattice sites). The degeneracy of the 
lowest eigenstate of $H_e$ is then given by
$$\frac{M!}{(\frac{M}{2})!}.$$

The degeneracy may be removed by
diagonalizing the first nontrivial order in perturbation theory. All
matrix elements in the vector space of degenerate vacua of the first
order Hamiltonian vanish, since $<0|H_q|0>=0$ (as $\int
dUU_{AB}^{(\dagger)}=0$ \cite{Creutz:ub}). The first nonvanishing 
contribution comes from second order.
At this order one should compute
\begin{equation}
\label{L2}
E^{(2)}_0=<0|H_m|0>+<0|\tilde{H}_q\frac{\Pi_0}{E_0-H_e}\tilde{H}_q|0>,
\end{equation}
\noindent where
$$<,>=\left\{\prod_{[\vec{r},\hat{n}]}\int dU[\vec{r},\hat{n}]\right\}(,)$$
\noindent is the inner product in the full Hilbert space of the model.
$dU$ is the Haar measure on the gauge group manifold and $(,)$
the fermion Fock space inner product;
$\Pi_0$ is the projection operator projecting onto states
orthogonal to $|0>$. In Eq.(\ref{L2}) $\Pi_0$ is ineffective since
the states created when $\tilde{H}_q$ acts on the vacuum 
are always orthogonal to $|0>$.  

$E^{(2)}_0$ may be more conveniently computed 
by constructing an eigenstate of $H_e$ and using it to evaluate the
function $f(H_e)=\frac{\Pi_0}{E_0-H_e}$ 
appearing in Eq.(\ref{L2}). Since the vacuum state
$|0>$ is a singlet of the electric field algebra, one has
\begin{equation}
E^a[\vec{r},\hat{n}]|0>=0
\end{equation}
which, in turn, implies that
\begin{equation}
\label{M}
H_e|0>=0.
\end{equation}
\noindent Using then Eq.(\ref{F}) and Eq.(\ref{M}) and putting the
commutator $[H_e,U[\vec{r},\hat{n}]]|0>$ in place of
$H_eU[\vec{r},\hat{n}]|0>$, one finds
\begin{equation}
\label{N}
H_eU[\vec{r},\hat{n}]|0>=\frac{g^2}{2a}C_2(N_C)U[\vec{r},\hat{n}]|0>,
\end{equation}
\noindent where $C_2=(N_C^2-1)/2N_C$ is the Casimir operator of
$SU(N_C)$. $U[\vec{r},\hat{n}]|0>$ is then an eigenstate of $H_e$ with
eigenvalue $g^2C_2(N_C)/2a$. Consequently,
\be
\label{fhe}
<0|\tilde{H}_q\frac{1}{E_0-H_e}\tilde{H}_q|0>=
-\frac{4a}{g^2C_2}<0|{H}^\dagger_q{H}_q|0>.
\ee
After integration over the link variables
$U$~\cite{Creutz:ub} [see also Eqs.(\ref{1U}),(\ref{2U})] 
one sees immediately that the first term in Eq.(\ref{L2}) is zero
and that the second order correction to the vacuum energy is given by
\begin{equation}
\label{O}
E^{(2)}_0=-\frac{1}{g^2aC_2N_C}<0|\sum_{[\vec{r},\hat{n}]}
\left[\rho(\vec{r}+\hat{n})+
\frac{N_C}{2}\right]\left[-\rho(\vec{r})+\frac{N_C}{2}\right]|0>.
\end{equation}
\noindent 
The diagonalized effective Hamiltonian is, up to an additive constant, given by
\begin{equation}
\label{Heff}
H_{\rm eff}=\frac{1}{g^2aC_2N_C}\sum_{[\vec{r},\hat{n}]}
\left[\rho(\vec{r}+\hat{n})\rho(\vec{r})\right].
\end{equation}
In deriving Eq.(\ref{Heff}) one should also take into account the fact we 
that gauge invariant
states such as $|0>$ should be charge neutral, 
i.e., $\sum_{\vec{r}}\rho(\vec{r})=0$.
As it is well known, $H_{eff}$ takes the form of the Hamiltonian of an 
antiferromagnet of spin $N_C/2$. It was shown in~\cite{Brower:1999ak}
that even for domain-wall fermions the effective Hamiltonian 
is that of an antiferromagnet but the fermions are massive and doubled.

Since, due to Eq.(\ref{rhoop}), $\rho(\vec r)$ has only two possible 
eigenvalues $\rho=\pm N_C/2$, the lowest
energy a link can have may occur only when one end has 
$\rho=+ N_C/2$ and the other has $\rho=-N_C/2$.
In the space of pure fermion states the true ground state must thus minimize 
$H_{\rm eff}$ and be fluxless. 

There are two ground states of $H_{\rm eff}$~\cite{Banks:1976ia} corresponding
to those of the antiferromagnetic Ising model of spin $N_C/2$.
In the large $N_C$ limit the two ground states are not mixed at
any finite order of perturbation theory. In fact, it would be necessary
to apply $H_q$ at least $N_C$ times in order to transform one ground state 
into the other, since
$H_q$ acts as an hopping Hamiltonian destroying a quark
on a site and creating it on a neighboring site.
One may choose as the ground state the one in which $\rho=+N_C/2$ 
on even sites and $\rho=-N_C/2$ on odd 
sites, the other state being obtained by interchanging odd and even sites.
With this choice of the ground state the sum over the lattice links 
$[\vec r,\hat n]$ in Eq.(\ref{O}) may be easily done and, for the
ground state energy at the second order in the strong-coupling 
perturbative expansion, one gets
\begin{equation}
\label{E02}
E_{0}^{(2)}=-\frac{N_C}{2g^2aC_2}N.
\end{equation}
As evidenced in Eq.(\ref{transl}), the chiral symmetry on the lattice 
is given by the translation by a 
single link and takes even sites into odd sites.
Having chosen one of the two ground states described above,
chiral symmetry is spontaneously broken~\cite{Banks:1976ia} 
in the large $N_C$ limit. Thus, in the perturbative
expansion one has to consider only diagonal matrix elements and,
consequently, perturbation theory for nondegenerate states. 

The ground state energy up to the fourth order in the 
strong coupling expansion, is given by
\begin{equation}
\label{P}
E_0=E_0^{(0)}+E^{(2)}_0+E^{(4)}_0,
\end{equation}
\noindent where
\begin{equation} 
E^{(4)}_0=[E^{(4)}_0]_I+[E^{(4)}_0]_{II}+[E^{(4)}_0]_{III}
\end{equation}
with
\begin{eqnarray}
\label{Q}
&&[E^{(4)}_0]_I=<0|\tilde{H}_q\frac{\Pi_0}{E_0^{(0)}-H_e}\tilde{H}_q\frac{\Pi_0}{E_0^{(0)}-H_e}
\tilde{H}_q\frac{\Pi_0}{E_0^{(0)}-H_e}\tilde{H}_q|0>\\
\label{Q1}
&&[E^{(4)}_0]_{II}=-<0|\tilde{H}_q\frac{\Pi_0}{E_0^{(0)}-H_e}\tilde{H}_q|0>
<0|\tilde{H}_q\frac{\Pi_0}{(E_0^{(0)}-H_e)^2}\tilde{H}_q|0>\\
\label{Q2}
&&[E^{(4)}_0]_{III}=<0|H_m\frac{\Pi_0}{E^{(0)}_0-H_e}H_m|0>.
\end{eqnarray}
Equation (\ref{Q1}) has the same form of a second order contribution and 
may be evaluated following the same steps used to arrive at
Eq.(\ref{E02}): the only difference being that, in Eq.(\ref{Q1}), 
the energy denominator also appears squared so that, using Eq.(\ref{N}), 
one gets
\be
<0|\tilde{H}_q\frac{1}{(E_0-H_e)^2}\tilde{H}_q|0>=
\frac{8a^2}{g^4C^2_2}<0|{H}^\dagger_q{H}_q|0>.
\ee
After integration over the link variables, one finds that
\begin{eqnarray}
\label{IVordv1}
&& <0|\tilde{H}_q\frac{\Pi_0}{E_0^{(0)}-H_e}\tilde{H}_q|0>
<0|\tilde{H}_q\frac{\Pi_0}{(E_0^{(0)}-H_e)^2}\tilde{H}_q|0>=\cr
&& -\frac{2}{g^6aC_2^3N_C^2}\left[<0|\sum_{[\vec r,\hat n]}n(\vec r+\hat n)
(n(\vec r)-N_C)|0>\right]^2=-\frac{N_C^2N^2}{2g^6aC_2^3},
\end{eqnarray}
where
\begin{equation}
n(\vec r)=\rho(\vec r)+\frac{N_C}{2}
\end{equation}
yields the number of fermions at the site $\vec r$.

One may now turn to the first term contributing to $E^{(4)}_0$, Eq.(\ref{Q}). 
Using Eq.(\ref{N}), one gets
\begin{eqnarray}
\label{4ordv}
&&<0|\tilde{H}_q\frac{\Pi_0}{E_0^{(0)}-H_e}\tilde{H}_q
\frac{\Pi_0}{E_0^{(0)}-H_e}
\tilde{H}_q\frac{\Pi_0}{E_0^{(0)}-H_e}\tilde{H}_q|0>=\\
\nonumber
&& \frac{4a^2}{g^4C_2^2}\left(4<0|H_qH_q^{\dagger}\frac{\Pi_0}{E_0^{(0)}-H_e}
H_qH_q^{\dagger}|0>+
2<0|H_qH_q\frac{\Pi_0}{E_0^{(0)}-H_e}H_q^{\dagger}H_q^{\dagger}|0>\right).
\end{eqnarray}
In deriving Eq.(\ref{4ordv}) from Eq.(\ref{Q}), one should observe 
that the most external $\Pi_0$ is ineffective;
the only task to accomplish is then to evaluate the energy denominator 
in the middle.
For this purpose it is most convenient to rewrite $H_qH_q^{\dagger}|0>$ and
$H_q^{\dagger}H_q^{\dagger}|0>$ as linear combinations of 
eigenvectors of $H_e$, and in order to do this
one should first consider the action of $H_e$ on $UU^\dagger$
at different links.
Using the commutators (\ref{F}), (\ref{F1}) and the fact that $H_e|0>=0$ 
one finds
\begin{eqnarray}
\label{R}
H_eU_{AB}[\vec{r},\hat{n}]U^{\dagger}_{CD}[\vec{r^{\prime}},\hat{m}]|0>&=&
\frac{g^2}{a}\left(C_2(N_C)U_{AB}[\vec{r},\hat{n}]
U^{\dagger}_{CD}[\vec{r^{\prime}},\hat{m}]\right.\cr
&+&\left.\frac{1}{2N_C}U_{AB}[\vec{r},\hat{n}]U_{CD}^{\dagger}[\vec{r},\hat{n}]
\delta([\vec{r},\hat{n}]-[\vec{r^{\prime}},\hat{m}])\right.\cr
&-&\left.\frac{1}{2}\delta_{AD}\delta_{BC}\delta([\vec{r},\hat{n}]-[\vec{r^{\prime}},
\hat{m}])\right)|0>.
\end{eqnarray}
\noindent
Keeping into account the action of $\Pi_0$
it is easy to see that 
$U_{AB}[\vec{r},\hat{n}]U^{\dagger}_{CD}[\vec{r^{\prime}},\hat{m}]|0>$
is an eigenstate of $f(H_e)=\Pi_0/(E_0^{(0)}-H_e)$ with eigenvalue 
\begin{equation}
\label{R1}
f(H_e)U_{AB}[\vec{r},\hat{n}]U^{\dagger}_{CD}[\vec{r^{\prime}},\hat{m}]|0>=
-\frac{a}{g^2C_2}U_{AB}[\vec{r},\hat{n}]
U^{\dagger}_{CD}[\vec{r^{\prime}},\hat{m}]|0>.
\end{equation}
A similar procedure shows that when $H_e$ acts on two $U$'s 
at different links, one gets
\begin{eqnarray}
\label{S}
H_eU_{AB}^{\dagger}[\vec{r},\hat{n}]U^{\dagger}_{CD}
[\vec{r^{\prime}},\hat{m}]|0>
&=&\frac{g^2}{a}\left(C_2(N_C)U_{AB}^{\dagger}[\vec{r},\hat{n}]
U^{\dagger}_{CD}[\vec{r^{\prime}},\hat{m}]\right.\cr
&-&\left.\frac{N_C+1}{2N_C}U_{AB}^{\dagger}[\vec{r},\hat{n}]
U^{\dagger}_{CD}[\vec{r},\hat{n}]
\delta([\vec{r},\hat{n}]-[\vec{r^{\prime}},\hat{m}])\right)|0>.
\end{eqnarray}
In Appendix B we show that the combination 
$\left(U_{AB}^{\dagger}[\vec{r},\hat{n}]U^{\dagger}_{CD}
[\vec{r^{\prime}},\hat{m}]-
U_{AB}^{\dagger}[\vec{r},\hat{n}]U^{\dagger}_{CD}[\vec{r},\hat{n}]\right)|0>$
is indeed an eigenstate of $H_e$ with eigenvalue $g^2C_2(N_C)/a$. 
Using this result and the
fact that $U_{AB}^{\dagger}[\vec{r},\hat{n}]U^{\dagger}_{CD}
[\vec{r},\hat{n}]|0>$ is an eigenstate of $H_e$ one can easily show that
\begin{eqnarray}
\label{S1}
&&f(H_e)U_{AB}^{\dagger}[\vec{r},\hat{n}]U^{\dagger}_{CD}
[\vec{r^{\prime}},\hat{m}]|0>
=-\frac{a}{g^2C_2}\left(U_{AB}^{\dagger}[\vec{r},\hat{n}]
U^{\dagger}_{CD}[\vec{r^{\prime}},\hat{m}]\right.\cr
&&\left.+\frac{1}{N_C-2}U_{AB}^{\dagger}[\vec{r},\hat{n}]
U^{\dagger}_{CD}[\vec{r},\hat{n}]\delta([\vec{r},\hat{n}]
-[\vec{r^{\prime}},\hat{m}])\right)|0>.
\end{eqnarray}
Taking into account Eq.(\ref{R1}) and Eq.(\ref{S1}), Eq.(\ref{4ordv}) 
becomes
\begin{eqnarray}
&& <0|\tilde{H}_q\frac{\Pi_0}{E_0^{(0)}-H_e}\tilde{H}_q
\frac{\Pi_0}{E_0^{(0)}-H_e}
\tilde{H}_q\frac{\Pi_0}{E_0^{(0)}-H_e}\tilde{H}_q|0>=\cr
&& \frac{4a^3}{g^6C_2^3}\left(-4<0|H_qH_q^{\dagger}H_q
H_q^{\dagger}|0>-2<0|H_qH_qH_q^{\dagger}
H_q^{\dagger}|0>\right.\cr                  
&& \left.-\frac{1}{2a^2(N_C-2)}<0|H_qH_q\sum_{[\vec{r},\hat{n}]}
\Psi_{A}^{\dagger}(\vec{r})U_{AB}^{\dagger}[\vec{r},\hat{n}]
\Psi_{B}(\vec{r}+\hat{n})\right.\cr
&&\left.\Psi_{C}^{\dagger}(\vec{r})
U_{CD}^{\dagger}[\vec{r},\hat{n}]
\Psi_{D}(\vec{r}+\hat{n})|0>\right).
\end{eqnarray}
Integrating now over the link variables (see Eqs.(\ref{2U}),(\ref{4U})), 
one gets 
\begin{eqnarray}
\label{IVordv2}
&& <0|\tilde{H}_q\frac{\Pi_0}{E_0^{(0)}-H_e}\tilde{H}_q
\frac{\Pi_0}{E_0^{(0)}-H_e}
\tilde{H}_q\frac{\Pi_0}{E_0^{(0)}-H_e}\tilde{H}_q|0>=\cr
&& -\frac{1}{g^6aC_2^3}\left[\frac{1}{N_C^2}<0|
\left(-\sum_{[\vec{r},\hat{n}]\neq[\vec{r}+\hat{n},\hat{m}]}
n(\vec{r}+\hat{n})(n(\vec{r})-N_C)(n(\vec{r}+\hat{n}+\hat{m})-N_C)
\right.\right.\cr
&& \left.\left.+\sum_{[\vec{r},\hat{n}]\neq[\vec{r}-\hat{m},\hat{m}]}
n(\vec{r}+\hat{n})(n(\vec{r})-N_C)n(\vec{r}-\hat{m})+\sum_{[\vec{r},\hat{n}]
\neq[\vec{r},\hat{m}]}n(\vec{r}+\hat{n})(n(\vec{r})-N_C)n(\vec{r}+\hat{m})
\right.\right.\cr
&& \left.\left.-\sum_{[\vec{r},\hat{n}]\neq[\vec{r}+\hat{n}-\hat{m},\hat{m}]}
n(\vec{r}+\hat{n})(n(\vec{r})-N_C)(n(\vec{r}+\hat{n}-\hat{m})-N_C)
\right.\right.\cr
&& \left.\left.+\sum_{[\vec{r},\hat{n}]\neq[\vec{r^{\prime}},\hat{m}]}
n(\vec{r}+\hat{n})(n(\vec{r})-N_C)n(\vec{r^{\prime}}+\hat{m})
(n(\vec{r^\prime)}-N_C)\right.\right.\cr
&& \left.\left.+\sum_{[\vec{r},\hat{n}]\neq[\vec{r^{\prime}},\hat{m}]}
n(\vec{r}+\hat{n})(n(\vec{r})-N_C)n(\vec{r^{\prime}})
(n(\vec{r^{\prime}}+\hat m)-N_C)\right)|0>\right.\cr
&& \left.+\frac{1}{N_C(N_C-2)}<0|\sum_{[\vec r,\hat n]}
\left(-n(\vec r+\hat n)(n(\vec r)-N_C)+
n(\vec r+\hat n)^2(n(\vec r)-N_C)\right.\right.\cr
&& \left.\left.-n(\vec r+\hat n)(n(\vec r)-N_C)^2+
n(\vec r+\hat n)^2(n(\vec r)-N_C)^2\right)|0>\right]=\cr
&& \frac{1}{2g^6aC_2^3}\left[-N_C^2N^2+\frac{N_C(10N_C-21)}{N_C-2}N\right].
\end{eqnarray}
In deriving Eq.(\ref{IVordv2}) one should use Eq.(\ref{4U}) 
when two $U$'s and two $U^\dagger$'s are on the same link.

Equation (\ref{Q2}) in $E_0^{(4)}$ is the magnetic contribution 
to the vacuum energy.
Since $H_m$ contains already a factor $1/g^2$,
it contributes only at the fourth order of the strong coupling expansion. 
It can be written as
\begin{eqnarray}
\label{Hm}
&& <0|H_m\frac{\Pi_0}{E_0^{(0)}-H_e}H_m|0>=\cr
&& 4<0|\sum_{[\vec r,\hat n,\hat m]}\frac{1}{2g^2a}U_{AB}[\vec r,\hat n]
U_{BC}[\vec r+\hat n,\hat m]U_{CD}^{\dagger}[\vec r+\hat m,\hat n]
U_{DA}^{\dagger}[\vec r,\hat m]\frac{\Pi_0}{E_0^{(0)}-H_e}\cr
&& \sum_{[\vec{r^{\prime}},\hat l,\hat k]}\frac{1}{2g^2a}
U_{EF}[\vec{r^{\prime}},
\hat l]U_{FG}[\vec{r^{\prime}}+\hat l,\hat k]U_{GH}^{\dagger}[\vec{r^{\prime}}
+\hat k,\hat l]U_{HE}^{\dagger}[\vec{r^{\prime}},\hat k]|0>.
\end{eqnarray}

Using Eq.(\ref{F}) and Eq.(\ref{F1}), one finds 
\begin{eqnarray}
&& H_eU_{AB}[\vec r,\hat n]
U_{BC}[\vec r+\hat n,\hat m]U_{CD}^{\dagger}[\vec r+\hat m,\hat n]
U_{DA}^{\dagger}[\vec r,\hat m]|0>=\cr
&& 2\frac{g^2}{a}C_2U_{AB}[\vec r,\hat n]
U_{BC}[\vec r+\hat n,\hat m]U_{CD}^{\dagger}[\vec r+\hat m,\hat n]
U_{DA}^{\dagger}[\vec r,\hat m].
\end{eqnarray}
Thus, a plaquette acting on the vacuum $|0>$, $Tr[UUU^\dagger U^\dagger]|0>$,
is an eigenstate of $H_e$ with eigenvalue $2g^2C_2/a$.
Equation (\ref{Hm}) then becomes
\bea
&& <0|H_m\frac{\Pi_0}{E_0^{(0)}-H_e}H_m|0>=\cr
&& -\frac{1}{2g^6aC_2}<0|\sum_{[\vec r,\hat n,\hat m]}U_{AB}[\vec r,\hat n]
U_{BC}[\vec r+\hat n,\hat m]U_{CD}^{\dagger}[\vec r+\hat m,\hat n]
U_{DA}^{\dagger}[\vec r,\hat m]\cr
&& \sum_{[\vec{r^{\prime}},\hat l,\hat k]}U_{EF}[\vec{r^{\prime}},
\hat l]U_{FG}[\vec{r^{\prime}}+\hat l,\hat k]U_{GH}^{\dagger}[\vec{r^{\prime}}
+\hat k,\hat l]U_{HE}^{\dagger}[\vec{r^{\prime}},\hat k]|0>.
\end{eqnarray}
The integration over the link variables requires that the 
two plaquettes are coincident, otherwise the integral would vanish. Keeping 
into account that $2N$ is the number of oriented plaquettes on the lattice, 
one gets
\be
\label{IVordv3}
<0|H_m\frac{\Pi_0}{E_0^{(0)}-H_e}H_m|0>=-\frac{N}{g^6aC_2}.
\ee

Collecting all the terms (\ref{IVordv1}), (\ref{IVordv2}) and (\ref{IVordv3})
the fourth order correction to the vacuum energy becomes
\begin{equation}
\label{T}
E^{(4)}_0=\frac{N}{2g^6aC_2^3}\frac{N_C(10N_C-21)}{N_C-2}
-\frac{N}{g^6aC_2}\equiv E^{(4)}_{q0}+E^{(4)}_{m0},
\end{equation}
where $E^{(4)}_{q0}$ is the contribution due to the quark Hamiltonian
and $E^{(4)}_{m0}$ the one due to the magnetic Hamiltonian. 
Note that $E_0^{(4)}$ 
is proportional to $N$ and thus the vacuum energy is an extensive variable.
Moreover, the $N^2$ dependence of Eq.(\ref{IVordv1}) is precisely canceled 
by Eq.(\ref{IVordv2}). As  a check of our computation one may set $N_C=3$ in 
Eq.(\ref{T}) and compare the result with those obtained in
~\cite{Banks:1976ia} using a completely different approach.
The agreement is exact
when the coefficient of the irrelevant operator introduced
in \cite{Banks:1976ia} is set to zero \footnote[1]{There should be a minor 
misprint in the final equation for $\omega_0^{(4)}$ in~\cite{Banks:1976ia}.
In fact if one adds up all the contributions from each graph
the result is slightly different and coincides with Eq.(\ref{T})
for $N_C=3$.}.

\section{The low-mass meson states}

In the strong coupling expansion the lowest-lying states in the meson
spectrum are those consisting of a quark and an antiquark at opposite
ends of a single link. If the quark is at $(\vec{r}+\hat{n})$ and the
antiquark at $\vec{r}$ a basis for such states is given by
\be
|\vec{r},\hat{n}>=\Psi^{\dagger}(\vec{r}+\hat{n})U[\vec{r},\hat{n}]
\Psi(\vec{r})|0>.
\label{basis}
\ee
For a given meson, the wave function may be determined through the following
steps.
One may first take the quark bilinear in the continuum with the desired 
transformation properties and appropriate continuum quantum numbers and 
then write it in point-separated lattice form using the discrete
symmetries of the theory. Only after fixing the pertinent lattice quantum 
numbers one may apply the bilinear to the vacuum. 

In the following we shall be interested only in
the low-lying unflavored mesons:
$\pi_0$, $\rho$, $\omega$, $b_1$, $a_1$, $f_2$,$f_0$. 
In the continuum theory the wave functions for these mesons are
given by
\begin{eqnarray}
\label{C1}
&& |\pi_0>\sim i\overline{\Psi}\gamma_5\frac{1}{2}\tau_3\Psi|0>\\
\label{C2}
&& |\omega>\sim \Psi^{\dagger}\alpha_z\Psi|0>\\
\label{C3}
&& |\rho>\sim \Psi^{\dagger}\alpha_x\frac{1}{2}\tau_3\Psi|0>\\
\label{C4}
&& |b_1>\sim i \Psi^{\dagger}\gamma_5\partial_z\tau_3\Psi|0>\\
\label{C5}
&& |a_1>\sim
i\Psi^{\dagger}(\alpha_x\partial_y-\alpha_y\partial_x)\tau_3\Psi|0>\\
\label{C6}
&& |f_2>\sim
i\Psi^{\dagger}(\alpha_z\partial_z+\alpha_x\partial_x-2\alpha_y\partial_y)\Psi|0>\\
\label{C7}
&& |f_0>\sim i\overline{\Psi}\nablaslash\Psi|0>.
\end{eqnarray}
The choice done is as in~\cite{Banks:1976ia} and it is based
on the quantum numbers labeling the mesonic states. For example, for $\pi_0$
one needs a pseudoscalar with nontrivial isospin; thus 
$i\overline{\Psi}\gamma_5\tau_3\frac{1}{2}\Psi$ is a pertinent choice
of the wave function. The choice 
for the components of the vector mesons ($\omega$, $\rho$,
$b_1$, $a_1$) or of the spin-2 meson ($f_2$) is made
by observing that these are the only
components of these mesons that on the lattice have the standard form 
(\ref{basis}).

The lattice form of these operators may be obtained by applying
the staggered fermion formalism to derive
operators with appropriate lattice quantum numbers. 
The lattice wave functions at zero momentum are then given by
\begin{eqnarray}
\label{M1}
 |\pi_0>&=&\frac{i}{\sqrt{N_CM}}\left[\sum_{\vec{r}}(-1)^{x}
\Psi_{A}^{\dagger}(\vec{r}+\hat{z})
U_{AB}[\vec{r},\hat{z}]\Psi_{B}(\vec{r})-h.c.\right]|0>\\
\label{M2}
|\omega>&=&\frac{i}{\sqrt{N_CM}}\left[\sum_{\vec{r}}(-1)^{y}
\Psi_{A}^{\dagger}(\vec{r}+\hat{z})
U_{AB}[\vec{r},\hat{z}]\Psi_{B}(\vec{r})-h.c.\right]|0>\\
\label{M3}
 |\rho>&=&\frac{1}{\sqrt{N_CM}}\left[\sum_{\vec{r}}(-1)^{y}
\Psi_{A}^{\dagger}(\vec{r}+\hat{y})
U_{AB}[\vec{r},\hat{y}]\Psi_{B}(\vec{r})+h.c.\right]|0>\\
\label{M4}
|b_1>&=&\frac{1}{\sqrt{N_CM}}\left[\sum_{\vec{r}}(-1)^{x}
\Psi_{A}^{\dagger}(\vec{r}+\hat{z})
U_{AB}[\vec{r},\hat{z}]\Psi_{B}(\vec{r})+h.c.\right]|0>\\
\label{M5}
|a_1>&=&\frac{i}{\sqrt{2N_CM}}\left[\sum_{\vec{r}}[(-1)^{y}
\Psi_{A}^{\dagger}(\vec{r}+\hat{y})
U_{AB}[\vec{r},\hat{y}]\Psi_{B}(\vec{r})\right.\\
\nonumber &
+&\left.(-1)^{x+y+z}\Psi_{A}^{\dagger}(\vec{r}+\hat{x})
U_{AB}[\vec{r},\hat{x}]\Psi_{B}(\vec{r})]-h.c.\right]|0>\\
\label{M6}
|f_2>&=&\frac{1}{\sqrt{2N_CM}}\left[\sum_{\vec{r}}[(-1)^{y}
\Psi_{A}^{\dagger}(\vec{r}+\hat{z})
U_{AB}[\vec{r},\hat{z}]\Psi_{B}(\vec{r})\right.\\
 \nonumber
&+&\left.(-1)^{z}\Psi_{A}^{\dagger}(\vec{r}+\hat{x})
U_{AB}[\vec{r},\hat{x}]\Psi_{B}(\vec{r})]+h.c.\right]|0>\\
\label{M7}
|f_0>&=&\frac{1}{\sqrt{3N_CM}}\left[\sum_{\vec{r}}\eta(\hat{n})
\Psi_{A}^{\dagger}(\vec{r}+\hat{n})U_{AB}[\vec{r},\hat{n}]
\Psi_{B}(\vec{r})+h.c.\right]|0>.
\end{eqnarray}  
The normalizations are fixed in the standard way by integrating
over the link variables.

All the mesons are degenerate at the lowest order and their
energy is given by
\be
\label{E0M}
E_M^{(0)}=<\mathcal{M}|H_e|\mathcal{M}>=\frac{g^2C_2}{2a},
\ee 
as it can be easily seen using Eq.(\ref{N}) and integrating over 
the link variables using Eq.(\ref{2U}).
Since a static limit (fixed
and large $a$) is used -and thus the states cannot propagate 
in this approximation- all
the single-link mesons have the same mass regardless of the character
(e.g., $s$-wave or $p$-wave) of their continuum wave functions. 
As we shall see, the fourth order computation will cure 
this unphysical effect of the static lattice approximation. 

The meson energy has been computed up
to the fourth order in the perturbative expansion using a method
similar to the one used in the evaluation of the vacuum
energy. If one evaluates the meson energy up to the fourth order, one gets
$$E_M=\frac{g^2C_2}{2a}+E_M^{(2)}+E_M^{(4)},$$
\noindent where
\begin{equation}
E_M^{(2)}=<\mathcal{M}|\tilde{H}_q\frac{\Pi_M}{E_M^{(0)}-H_e}
\tilde{H}_q|\mathcal{M}>
\end{equation}
and
\begin{eqnarray}
\label{IV1}
E_M^{(4)}= &&
<\mathcal{M}|\tilde{H}_q\frac{\Pi_M}{E_M^{(0)}-H_e}\tilde{H}_q
\frac{\Pi_M}{E_M^{(0)}-H_e}\tilde{H}_q\frac{\Pi_M}{E_M^{(0)}-H_e}
\tilde{H}_q|\mathcal{M}>\\
\label{IV2}
&& -
<\mathcal{M}|\tilde{H}_q\frac{\Pi_M}{E_M^{(0)}-H_e}\tilde{H}_q|\mathcal{M}>
<\mathcal{M}|\tilde{H}_q\frac{\Pi_M}{(E_M^{(0)}-H_e)^2}
\tilde{H}_q|\mathcal{M}>\\
\label{IV3}
&& +<\mathcal{M}|H_m\frac{\Pi_M}{E_M^{(0)}-H_e}H_m|\mathcal{M}>.
\end{eqnarray}
$\Pi_M$ is the projection operator projecting onto states
orthogonal to those states that have the unperturbed energy of a 
generic single link meson (\ref{E0M}).
$|\mathcal{M}>$ is the meson operator.

\subsection{Second order}

The projection operator $\Pi_M$ does not affect the second order in the
strong coupling expansion since the states created by $\mathcal{M}$ and 
$\tilde{H}_q$  acting on the vacuum $|0>$ are orthogonal to the meson state.
Thus, the matrix elements to be computed are
\begin {equation}
\label{U}
E_{M}^{(2)}=<(M+M^{\dagger})|(H_q+H_q^{\dagger})
\frac{1}{E_M^{(0)}-H_e}(H_q+H_q^{\dagger})|(M+M^{\dagger})>,
\end{equation}
where $|\mathcal{M}>$ is one of the meson operator
(\ref{M1})-(\ref{M7}) and $\mathcal{M}\equiv M+M^{\dagger}$. 
In Eq.(\ref{U}) the terms containing a different number of $U$ and $U^\dagger$
vanish when integrated over the link variables.
The remaining terms may be conveniently grouped as 
\begin{equation}
\label{2nd}
E_M^{(2)}=2\left[<M|H_q\tilde\Lambda_{M}(H_e)H_q^{\dagger}|M^{\dagger}>+
<M|H_q^{\dagger}\tilde\Lambda_{M}(H_e)H_q|M^{\dagger}>+
<M|H_q^{\dagger}\tilde\Lambda_{M}(H_e)H_q^{\dagger}|M>\right],
\end{equation}
\noindent where
$$\tilde\Lambda_{M}(H_e)=\frac{1}{E_{M}^{(0)}-H_e}.$$
\noindent All the one-link meson states (\ref{M1})-(\ref{M7}) consist
of linear combinations of directed links on the
lattice with appropriate phases, which are responsible
for the differences in the matrix elements of the various mesons 
(\ref{M1})-(\ref{M7}).

To elucidate the method used to compute these matrix elements, 
it is sufficient to consider a generic meson of the 
form
\be
\sum_{\vec{r}}S(\vec{r})
\Psi_{A}^{\dagger}(\vec{r}+\hat{n})U_{AB}[\vec{r},\hat{n}]\Psi_{B}(\vec{r})|0>;
\label{gm}
\ee
In Eq.(\ref{gm}) $S(\vec{r})$ is one of the phases of the meson operators
(\ref{M1})-(\ref{M7}).
In order to compute
\begin{eqnarray}
\label{II1}
&&(II)_1=<M|H_q\tilde\Lambda_{M}(H_e)H_q^{\dagger}|M^{\dagger}>\\
\label{II2}
&&(II)_2=<M|H_q^{\dagger}\tilde\Lambda_{M}(H_e)H_q|M^{\dagger}>\\
\label{II3}
&&(II)_3=<M|H_q^{\dagger}\tilde\Lambda_{M}(H_e)H_q^{\dagger}|M>
\end{eqnarray}
one should construct suitable eigenstates of the unperturbed
Hamiltonian $H_e$ in order to evaluate the function of $H_e$,
$\tilde\Lambda_{M}(H_e)$. Using the 
eigenvectors found in the previous section, it is quite easy 
to obtain
\begin{eqnarray}
\label{U1}
&&\tilde\Lambda_{M}(H_e)U_{AB}^{\dagger}[\vec{r},\hat{n}]
U_{CD}^{\dagger}[\vec{r^{\prime}},
\hat{m}]|0>=\cr
&&\frac{2a}{g^2C_2}\left(-U_{AB}^{\dagger}[\vec{r},\hat{n}]U_{CD}^{\dagger}
[\vec{r^{\prime}},\hat{m}]-\frac{2}{N_C-3}U_{AB}^{\dagger}[\vec{r},\hat{n}]
U_{CD}^{\dagger}[\vec{r},\hat{n}]\right)|0>
\end{eqnarray}
\noindent and
\begin{eqnarray}
\label{U2}
&&\tilde\Lambda_{M}(H_e)U_{AB}[\vec{r},\hat{n}]
U_{CD}^{\dagger}[\vec{r^{\prime}},\hat{m}]|0>=\cr
&&\frac{2a}{g^2C_2}\left(-U_{AB}[\vec{r},\hat{n}]U_{CD}^{\dagger}
[\vec{r^{\prime}},\hat{m}]+\frac{2}{N_C^2+1}U_{AB}[\vec{r},\hat{n}]
U_{CD}^{\dagger}[\vec{r},\hat{n}]\delta\left([\vec{r},\hat{n}]-
[\vec{r^{\prime}},\hat{m}]\right)\right.\cr
&&\left.+\frac{2N_C}{N_C^2+1}\delta_{AD}\delta_{BC}
\delta\left([\vec{r},\hat{n}]-[\vec{r^{\prime}},\hat{m}]\right)\right)|0>.
\end{eqnarray}
\noindent Plugging Eq.(\ref{U1}) and Eq.(\ref{U2}) in
Eqs.(\ref{II1})-(\ref{II3}), one gets
\begin{eqnarray}
\label{II1a}
(II)_1=&-&\frac{2a}{g^2C_2}<M|H_qH_q^{\dagger}|M^{\dagger}>\cr
&-&\frac{4}{g^2C_2(N_C-3)}<M|H_q\frac{1}{2}\sum_{[\vec{r},\hat{n}]}
\eta(\hat{n})\Psi_{A}^{\dagger}(\vec{r})U_{AB}^{\dagger}[\vec{r},\hat{n}]
\Psi_{B}(\vec{r}+\hat{n})\cr
&&\phantom{\frac{4}{g^2C_2(N_C-3)}}\sum_{\vec{r}}S^{\star}(\vec{r})
\Psi_{C}^{\dagger}(\vec{r})U_{CD}^{\dagger}[\vec{r},\hat{n}]
\Psi_{D}(\vec{r}+\hat{n})|0>.
\end{eqnarray}
\begin{eqnarray}
\label{II2a}
(II)_2=&-&\frac{2a}{g^2C_2}<M|H_q^{\dagger}H_q|M^{\dagger}>\cr
&+&\frac{4}{g^2C_2(N_C^2+1)}<M|H_q^{\dagger}
\frac{1}{2}\sum_{[\vec{r},\hat{n}]}\eta(\hat{n})
\Psi_{A}^{\dagger}(\vec{r}+\hat{n})U_{AB}[\vec{r},\hat{n}]\Psi_{B}(\vec{r})\cr
&&\phantom{\frac{4}{g^2C_2(N_C^2+1)}}
\sum_{\vec{r}}S^{\star}(\vec{r})\Psi_{C}^{\dagger}(\vec{r})
U_{CD}[\vec{r},\hat{n}]\Psi_{D}(\vec{r}+\hat{n})|0>\cr
&+&\frac{4N_C}{g^2C_2(N_C^2+1)}<M|H_q^{\dagger}\frac{1}{2}
\sum_{[\vec{r},\hat{n}]}\eta(\hat{n})\Psi_{A}^{\dagger}(\vec{r}+\hat{n})
\Psi_{B}(\vec{r})\cr
&&\phantom{\frac{4N_C}{g^2C_2(N_C^2+1)}}\sum_{r}S^{\star}(\vec{r})
\Psi_{C}^{\dagger}(\vec{r})
\Psi_{D}(\vec{r}+\hat{n})\delta_{AD}\delta_{BC}|0>.
\end{eqnarray}
\begin{eqnarray}
\label{II3a}
(II)_3=&-&\frac{2a}{g^2C_2}<M|H_q^{\dagger}H_q^{\dagger}|M>\cr
&+&\frac{4}{g^2C_2(N_C^2+1)}<M|H_q^{\dagger}\frac{1}{2}
\sum_{[\vec{r},\hat{n}]}\eta(\hat{n})\Psi_{A}^{\dagger}
(\vec{r})U_{AB}^{\dagger}[\vec{r},\hat{n}]\Psi_{B}(\vec{r}+\hat{n})\cr
&&\phantom{\frac{4}{g^2C_2(N_C^2+1)}}\sum_{\vec{r}}
S(\vec{r})\Psi_{C}^{\dagger}(\vec{r}+\hat{n})U_{CD}[\vec{r},\hat{n}]
\Psi_{D}(\vec{r})|0>\cr
&+&\frac{4N_C}{g^2C_2(N_C^2+1)}<M|H_q^{\dagger}\frac{1}{2}
\sum_{[\vec{r},\hat{n}]}\eta(\hat{n})\Psi_{A}^{\dagger}
(\vec{r})\Psi_{B}(\vec{r}+\hat{n})\cr
&&\phantom{\frac{4N_C}{g^2C_2(N_C^2+1)}}\sum_{r}S(\vec{r})
\Psi_{C}^{\dagger}(\vec{r}+\hat{n})\Psi_{D}(\vec{r})\delta_{AD}\delta_{BC}|0>.
\end{eqnarray}
One now needs the results of Eq.(\ref{2U}) and Eq.(\ref{4U}) to integrate over
the link variable $U$ and to choose the particular $S(\vec r)$
selecting a given meson.

As an example, consider 
the case of the $\rho$ meson. In terms of the fermion number
operator $n(\vec r)$ one has
\bea
&& [(II)_1]_{\rho}=-\frac{1}{2g^2aC_2N_C^2}<0|\left[\sum_{\vec r,
[\vec r,\hat y]\neq[\vec r,\hat m]}n(\vec r+\hat n)
(n(\vec r)-N_C)(n(\vec r+\hat m)\right.\cr
&& \left.-\sum_{\vec r,
[\vec r,\hat y]\neq[\vec r+\hat y-\hat m,\hat m]}n(\vec r+\hat y)
(n(\vec r)-N_C)(n(\vec r+\hat y-\hat m)\right.\cr
&& \left.+\sum_{\vec r,[\vec r,\hat y]\neq[\vec R,\hat m]}n(\vec r+\hat y)
(n(\vec r)-N_C)n(\vec R+\hat m)(n(\vec R)-N_C)\right.\cr
&& \left.+\sum_{\vec r,\vec R,[\vec r,\hat y]\neq[\vec R,\hat y]}
(-1)^{x+y+X+Y}n(\vec r+\hat y)(n(\vec r)-N_C)n(\vec R+\hat y)
(n(\vec R)-N_C)\right.\cr
&& \left.+\frac{2N_C}{N_C-3}\sum_{\vec r}\left(-n(\vec r+\hat y)(n(\vec r)-N_C)
-n(\vec r+\hat y)(n(\vec r)-N_C)^2\right.\right.\cr
&&\left.\left.+n(\vec r+\hat y)^2
(n(\vec r)-N_C)+n(\vec r+\hat y)^2(n(\vec r)-N_C)^2\right)\right]|0>\cr
&& =\frac{1}{2g^2aC_2}\left(3-\frac{2(N_C-1)}{N_C-3}
-\frac{N_C}{4}N\right).
\eea

\bea
&& [(II)_2]_{\rho}=-\frac{1}{2g^2aC_2N_C^2}<0|\left\{-\sum_{\vec r,
[\vec r,\hat y]\neq[\vec r+\hat y,\hat m]}n(\vec r+\hat y)
(n(\vec r)-N_C)(n(\vec r+\hat y+\hat m)-N_C)\right.\cr
&& \left.+\sum_{\vec r,[\vec r,\hat y]\neq[\vec r-\hat m,\hat m]}
n(\vec r+\hat y)(n(\vec r)-N_C)n(\vec r-\hat m)\right.\cr
&& \left.+\sum_{\vec r,[\vec r,\hat y]\neq[\vec R,\hat m]}
n(\vec r+\hat y)(n(\vec r)-N_C)n(\vec R)(n(\vec R+\hat m)-N_C)\right.\cr
&& \left.+\sum_{\vec r,\vec R,[\vec r,\hat y]\neq[\vec R,\hat y]}
(-1)^{x+y+X+Y}n(\vec r+\hat y)(n(\vec r)-N_C)n(\vec R+\hat y)
(n(\vec R)-N_C)\right.\cr
&& \left.+\sum_{\vec r}\left(\frac{N_C^2}{N_C^2-1}[n(\vec r+\hat y)^2
(n(\vec r)-N_C)^2-n(\vec r+\hat y)(n(\vec r)-N_C)]\right.\right.\cr
&& \left.\left.-\frac{N_C}{N_C^2-1}[-n(\vec r+\hat y)^2(n(\vec r)-N_C)
+n(\vec r+\hat y)(n(\vec r)-N_C)^2]\right)\right.\cr
&& \left.-\frac{2}{N_C^2+1}\left[\frac{N_C^2}{N_C^2-1}\left(\sum_{\vec r}
[n(\vec r+\hat y)^2
(n(\vec r)-N_C)^2-n(\vec r+\hat y)(n(\vec r)-N_C)]\right.\right.\right.\cr
&& \left.\left.\left.-\frac{N_C}{N_C^2-1}\sum_{\vec r}
[-n(\vec r+\hat y)^2(n(\vec r)-N_C)
+n(\vec r+\hat y)(n(\vec r)-N_C)^2]\right)\right.\right.\cr
&& \left.\left.+\sum_{\vec r,\vec R,[\vec r,\hat y]\neq[\vec R,\hat y]}
(-1)^{x+y+X+Y}n(\vec r+\hat y)(n(\vec r)-N_C)n(\vec R+\hat y)
(n(\vec R)-N_C)\right]\right.\cr
&& \left.-\frac{2N_C^2}{N_C^2+1}\sum_{\vec r,\vec R}(-1)^{x+y+X+Y}
n(\vec r+\hat y)(n(\vec r)-N_C)n(\vec R+\hat y)(n(\vec R)-N_C)\right\}|0>\cr
&& =\frac{1}{2g^2aC_2}\left(3-\frac{N_C}{4}N\right).
\eea

\bea
&& [(II)_3]_{\rho}=\frac{1}{2g^2aC_2N_C^2}<0|\left[
\sum_{\vec r}n(\vec r+\hat y)
(n(\vec r)-N_C)n(\vec r-\hat y)\right.\cr
&& \left.-\sum_{\vec r}n(\vec r+\hat y)(n(\vec r)-N_C)
(n(\vec r+2\hat y)-N_C)\right]|0>=-\frac{1}{2g^2aC_2}.
\eea
Adding up the three terms, one finds
\begin{equation}
\label{so3}
 E_{\rho}^{(2)}=\frac{1}{g^2aC_2}\left[5-\frac{2(N_C-1)}{N_C-3}
-\frac{N_C}{2}N\right].
\end{equation}

A similar procedure yields the following results for the second
order correction to the meson energies
\begin{eqnarray}
\label{so1}
&& E_{\pi_0}^{(2)}=\frac{1}{g^2aC_2}\left[5-\frac{2(N_C-1)}{N_C-3}
-\frac{N_C}{2}N\right]\\
\label{so2}
&& E_{\omega}^{(2)}=\frac{1}{g^2aC_2}\left[5-\frac{2(N_C-1)}{N_C-3}
-\frac{N_C}{2}N\right]\\
\label{so4}
&& E_{b_1}^{(2)}=\frac{1}{g^2aC_2}\left[7-\frac{2(N_C-1)}{N_C-3}
-\frac{N_C}{2}N\right]\\
\label{so5}
&& E_{a_1}^{(2)}=\frac{1}{g^2aC_2}\left[9-\frac{2(N_C-1)}{N_C-3}
-\frac{N_C}{2}N\right]\\
\label{so6}
&&E_{f_2}^{(2)}=\frac{1}{g^2aC_2}\left[9-\frac{2(N_C-1)}{N_C-3}
-\frac{N_C}{2}N\right]\\
\label{so7}
&& E_{f_0}^{(2)}=\frac{1}{g^2aC_2}\left[11-\frac{2(N_C-1)}{N_C-3}
-\frac{N_C}{2}N\right]
\end{eqnarray}
\noindent where the $N$-dependent terms cancel against the vacuum
energy, as should be since the masses are intensive quantities.\\ 
\indent 

After rescaling the coupling constant
according to the 't Hooft prescription (large $N_C$ with $g^2N_C$
fixed)
$$g^2N_C\rightarrow g^2$$
\noindent one finds that the large $N_C$ limit makes
Eqs.(\ref{so3})-(\ref{so7}) finite.

Our results have been checked
by deriving the second order meson masses for $N_C$ generic 
with the graphical procedure of Ref.\cite{Banks:1976ia}.
In this case also there is complete agreement.

\subsection{A comment on irrelevant operators}

Since the strong-coupling limit of QCD is not universal, adding
an irrelevant operator to the Hamiltonian, leads to the same 
physical predictions
in the continuum limit. This allows one to introduce
arbitrary parameters, the coefficients of these irrelevant operators,
which are then fixed by fitting the experimental data.
Our analysis shows that in the 
't Hooft limit ($N_C\to\infty$, $g^2N_C$ fixed) the meson masses can be made
independent of arbitrary parameters and that results in agreement 
with experiments
can be obtained without introducing irrelevant operators.

In the celebrated computation of the hadron spectrum by Banks et al.
\cite{Banks:1976ia} the lattice Hamiltonian was indeed modified by 
the addition of an irrelevant operator given by
\be
\label{irrop}
W=A\sum_{\vec r,\hat n}\left(\rho(\vec r)\rho(\vec r+\hat n)+
\frac{N_C^2}{4}\right)
\ee
where $A$ is a dimensionless irrelevant parameter.
The new term was chosen according to three demands.
First, it must remove the degeneracies at zeroth order so that 
nondegenerate perturbation theory can be used. Second, it must preserve the
symmetries of the original Hamiltonian.  The vacuum state of the modified 
theory must again break the chiral symmetry spontaneously. Third, the 
added term should have no effect on the continuum limit of the lattice 
theory, so it should be an irrelevant operator. 
$W$ is a four-fermion operator and, when it is written in terms 
of the continuum variables with the conventional units, it depends on
$Ag^2a^2$ so that in the continuum limit ($a\to 0$) it vanishes faster than 
$a^2$. 

The irrelevant operator (\ref{irrop}) was introduced by Banks et al.~\cite{Banks:1976ia}
mainly because in this way the meson masses are well defined even for $N_C=3$
at the second order in the strong coupling expansion.
The corrections (\ref{so3})-(\ref{so7}) are in fact 
divergent for $N_C=3$. This problem was avoided
by introducing $W$ in the
unperturbed Hamiltonian. 
The meson masses then depend on the arbitrary parameter $A$.
For $N_C=3$ the diverging contribution, evidenced in 
Eqs.(\ref{so3})-(\ref{so7}), comes from
a term in which the meson operators and the quark Hamiltonian are on 
the same link. In \cite{Banks:1976ia}
this term varies as $A^{-1}$ and if $A$ is set to zero it yields
a diverging contribution. It is not difficult to verify that,
if the irrelevant operator is not introduced,
up to the fourth order in the strong coupling expansion, there are 
divergences of the type $1/(N_C-1),\ 1/(N_C-2),\ 1/(N_C-3)$ and $1/(N_C-4)$;
at the next order there are divergences up to $1/(N_C-5)$; and so on.
In the infinite $N_C$ limit these divergences are avoided.
Our analysis thus shows that in the large $N_C$ approach to lattice QCD
there is no need for an irrelevant operator; 
in fact, with the 't Hooft prescription,
the limit $N_C\to \infty$ yields series expansions for the meson masses
which are free of divergences and thus well defined.

The constant $A$ in Eq.(\ref{irrop}) was fixed
in \cite{Banks:1976ia} by requiring 
that the $q\bar q$ state is less
massive than a nucleon-antinucleon state in the static limit. 
As we will show below, in the 't Hooft limit however, the baryon
masses are zero at zeroth order in the strong coupling expansion and 
acquire a mass proportional to $N_C$ only at second order. Thus, baryons
may be consistently regarded as QCD solitons
\cite{Witten:1979kh} and the unperturbed mass of a bound state $n\bar n$
in the large $N_C$ limit vanishes at the lowest order in the perturbative expansion.
A $q\bar q$ state is then not degenerate with a $n\bar n$ state
and for this reason also there is no need of an irrelevant operator.

Nucleon masses should be determined using an antisymmetric operator 
creating $N_C$ quarks at the same lattice site when acting on the vacuum state.
The normalized nucleon state is
\be
\label{opnucl}
|n>=\frac{1}{N_C!}\sqrt{\frac{2}{M}}\sum_{\vec r}\epsilon^{A_1 A_2\dots A_{N_C}}\Psi_{A_1}^{\dagger}(\vec r)
\Psi_{A_2}^{\dagger}(\vec r)\dots \Psi_{A_{N_C}}^{\dagger}(\vec r)|0>
\ee
At zeroth order in the strong-coupling expansion, the baryon is massless, 
since the creation operator $|n>$ does not contain any color flux and thus
$H_e|n>=0$. This is in agreement with the requirement that the nucleon mass,
being the mass of a soliton,
should vary as the inverse of the coupling constant.
At the second order in the strong coupling expansion baryons already acquire mass
given by
\bea
\label{En2}
&& E_n^{(2)}=<n|\tilde H_q\frac{\Pi_n}{E_n^{(0)}-H_e}\tilde H_q|n>=
\frac{1}{g^2aC_2}\left[-\frac{N_C}{2}N+N_C\right]\cr
&& m_n=E_n^{(2)}-E_0^{(2)}=\frac{1}{g^2aC_2}N_C
\eea
which, after rescaling the coupling constant according to the 
't Hooft prescription
$g^2N_C \to g^2$, varies as $N_C$ in the large $N_C$ limit.
$1/N_C$ is the ``coupling constant''; thus the baryon mass again varies
as the inverse of the coupling constant as a soliton mass should 
do~\cite{Witten:1979kh}.

\subsection{Fourth order}

The fourth order corrections to the meson energies are given by 
the matrix elements (\ref{IV1}), (\ref{IV2}) and
(\ref{IV3}), where the projection operator $\Pi_M$ eliminates the
states proportional to $|M>$. Again one can construct eigenstates of the
unperturbed Hamiltonian to evaluate the function of $H_e$ in
Eqs.(\ref{IV1}), (\ref{IV2}) and (\ref{IV3}). 

First consider Eq.(\ref{IV1}). The non vanishing terms in Eq.(\ref{IV1})
can be grouped as follows
\begin{eqnarray}
\label{7ter}
&&
<M|H_q\frac{\Pi_M}{E_M^{(0)}-H_e}H_q\frac{\Pi_M}{E_M^{(0)}-H_e}
H_q\frac{\Pi_M}{E_M^{(0)}-H_e}H_q|M>\cr
&&=2\left[<M^{\dagger}|H^{\dagger}_q\Lambda_{M}H_q^{\dagger}
\Lambda_{M}H_q\Lambda_{M}H_q|M>+
<M|H^{\dagger}_q\Lambda_{M}H_q^{\dagger}\Lambda_{M}H_q
\Lambda_{M}H_q|M^{\dagger}>\right.\cr
&&\left.+<M|H_q\Lambda_{M}H_q^{\dagger}\Lambda_{M}H_q
\Lambda_{M}H_q^{\dagger}|M^{\dagger}>
+<M|H^{\dagger}_q\Lambda_{M}H_q\Lambda_{M}H_q^{\dagger}
\Lambda_{M}H_q|M^{\dagger}>\right]\cr
&&+4\left[<M|H_q\Lambda_{M}H_q^{\dagger}\Lambda_{M}
H_q^{\dagger}\Lambda_{M}H_q|M^{\dagger}>
+<M|H_q\Lambda_{M}H_q^{\dagger}\Lambda_{M}H_q^{\dagger}
\Lambda_{M}H_q^{\dagger}|M>\right.\cr
&&\left.+<M|H_q^{\dagger}\Lambda_{M}H_q\Lambda_{M}
H_q^{\dagger}\Lambda_{M}H_q^{\dagger}|M>\right],
\end{eqnarray}
\noindent where
$$\Lambda_M(H_e)=\frac{\Pi_M}{E_M^{(0)}-H_e}.$$
To compute these matrix elements, first note
that the two external projection operators do not affect the
calculations. The projection operator in the middle, instead, does not allow 
patterns in which there are fermion operators creating 
and destroying the same quark at
the same lattice site. For example, a term of the form
$$...\frac{\Pi_M}{E_M^{(0)}-H_e}\frac{1}{4a^2}\sum_{[\vec{r},\hat{n}]}
\Psi_{A}^{\dagger}(\vec{r})U_{AB}^{\dagger}[\vec{r},\hat{n}]
\Psi_{B}(\vec{r}+\hat{n})\Psi_{C}^{\dagger}(\vec{r}+
\hat{n})U_{CD}[\vec{r},\hat{n}]\Psi_{D}(\vec{r})|M>$$
is eliminated by the projection operator since it gives rise to a state
with the same energy of a single link meson. 

In order to illustrate the method
used in the computation of the meson energy, one should focus
the attention on the first two terms in Eq.(\ref{7ter}) and evaluate
them for the generic meson (\ref{gm}). Note that in the first of the 
two terms the
projection operators are irrelevant. Let us define
\begin{equation}
\label{A7term}
(A)=<M^{\dagger}|H_q^{\dagger}\Lambda_{M}H_q^{\dagger}\Lambda_{M}
H_q\Lambda_{M}H_q|M>
\end{equation}
\noindent Using Eq.(\ref{U1}), (\ref{A7term}) may be rewritten as
\begin{eqnarray}
\label{IV1A}
&&(A)=\frac{4a^2}{g^4C_2^2}\left[<M^{\dagger}|H_q^{\dagger}H_q^{\dagger}
\Lambda_{M}H_qH_q|M>\right.\cr
&&\left.+\frac{2}{N_C-3}<M^{\dagger}|H_q^{\dagger}H_q^{\dagger}
\Lambda_{M}H_q\right.\cr
&&\left.\sum_{\vec r, [\vec{r},\hat{n}]}\eta(\hat{n})\Psi_{A}^{\dagger}
(\vec{r}+\hat{n})U_{AB}[\vec{r},\hat{n}]\Psi_{B}(\vec{r})
S(\vec{r})\Psi_{C}^{\dagger}(\vec{r}+\hat{n})U_{CD}[\vec{r},\hat{n}]
\Psi_{D}(\vec{r})|0>\right.\cr
&&\left.+\frac{1}{(N_C-3)^2}<0|\sum_{\vec{r},[\vec{r},\hat{n}]}S(\vec{r})
\Psi_{A}^{\dagger}(\vec{r})
U_{AB}^{\dagger}[\vec{r},\hat{n}]\Psi_{B}(\vec{r}+\hat{n})
\eta(\hat{n})\Psi_{C}^{\dagger}(\vec{r})U_{CD}^{\dagger}
[\vec{r},\hat{n}]\Psi_{D}(\vec{r}+\hat{n})
H_q^{\dagger}\Lambda_{M}H_q\right.\cr
&&\left.\sum_{\vec{r^{\prime}},[\vec{r^{\prime}},\hat{m}]}
\eta(\hat{m})\Psi_{E}^{\dagger}(\vec{r^{\prime}}+\hat{m})
U_{EF}[\vec{r^{\prime}},\hat{m}]
\Psi_{F}(\vec{r^{\prime}})S(\vec{r^{\prime}})
\Psi_{G}^{\dagger}(\vec{r^{\prime}}+\hat{m})U_{GH}[\vec{r^{\prime}},\hat{m}]
\Psi_{H}(\vec{r^{\prime}})|0>\right].
\end{eqnarray}
\noindent Making use of Eq.(\ref{F}), Eq.(\ref{F1}), Eq.(\ref{M})
and after constructing a suitable eigenstate of $H_e$, one finds
\begin{eqnarray}
\label{3U}
&&\Lambda_M(H_e)U_{AB}[\vec{r},\hat{n}]U_{CD}[\vec{r^{\prime}},
\hat{m}]U_{EF}[\vec{r^{\prime\prime}},\hat{l}]|0>\cr
&&=-\frac{a}{g^2C_2}\left[U_{AB}[\vec{r},\hat{n}]U_{CD}[\vec{r^{\prime}},
\hat{m}]U_{EF}[\vec{r^{\prime\prime}},\hat{l}]+\frac{1}{N_C-2}
\left(U_{AB}[\vec{r},\hat{n}]U_{CD}[\vec{r},\hat{n}]
U_{EF}[\vec{r^{\prime\prime}},\hat{l}]\right.\right.\cr
&&\left.\left.+U_{AB}[\vec{r},\hat{n}]
U_{CD}[\vec{r^{\prime}},\hat{m}]U_{EF}[\vec{r},\hat{n}]+U_{AB}[\vec{r},\hat{n}]
U_{CD}[\vec{r^{\prime}},\hat{m}]U_{EF}[\vec{r^{\prime}},\hat{m}]
\right)\right.\cr
&&\left.+\frac{6}{(N_C-4)(N_C-2)}U_{AB}[\vec{r},\hat{n}]
U_{CD}[\vec{r},\hat{n}]
U_{EF}[\vec{r},\hat{n}]\right]|0>.
\end{eqnarray}
\noindent Taking into account Eq.(\ref{3U}), Eq.(\ref{IV1A}) becomes
\begin{eqnarray}
\label{UUU}
&&(A)=\cr
&&\frac{4a^3}{g^6C_2^3}\left\{-<M^{\dagger}
|H_q^{\dagger}H_q^{\dagger}H_qH_q|M>\right.\cr
&&\left.-\frac{1}{N_C-2}\left[<M^{\dagger}|H_q^{\dagger}H_q^{\dagger}
\frac{1}{4a^2}\sum_{[\vec{r},\hat{n}]}\Psi_{G}^{\dagger}(\vec{r}
+\hat{n})U_{GH}[\vec{r},\hat{n}]\Psi_{H}(\vec{r})
\Psi_{I}^{\dagger}(\vec{r}+\hat{n})U_{IL}[\vec{r},\hat{n}]
\Psi_{L}(\vec{r})|M>\right.\right.\cr
&&\left.\left.+<M^{\dagger}
|H_q^{\dagger}H_q^{\dagger}\frac{1}{2a}\sum_{[\vec{r},\hat{n}]}\eta(\hat{n})
\Psi_{G}^{\dagger}(\vec{r}+\hat{n})U_{GH}[\vec{r},\hat{n}]
\Psi_{H}(\vec{r})H_q\sum_{\vec{r}}S(\vec{r})\Psi_{M}^{\dagger}(\vec{r}+\hat{n})
U_{MN}[\vec{r},\hat{n}]\Psi_{N}(\vec{r})|0>\right.\right.\cr
&&\left.\left.+<M^{\dagger}|H_q^{\dagger}
H_q^{\dagger}H_q\frac{1}{2a}\sum_{[\vec{r},\hat{n}]}\eta(\hat{n})
\Psi_{I}^{\dagger}(\vec{r}+\hat{n})
U_{IL}[\vec{r},\hat{n}]\Psi_{L}(\vec{r})\sum_{\vec{r}}S(\vec{r})
\Psi_{M}^{\dagger}(\vec{r}+\hat{n})
U_{MN}[\vec{r},\hat{n}]\Psi_{N}(\vec{r})|0>\right]\right.\cr
&&\left.-\frac{6}{(N_C-4)(N_C-2)}<M^{\dagger}|H_q^{\dagger}H_q^{\dagger}
\frac{1}{4a^2}\sum_{[\vec{r},\hat{n}]}\Psi_{G}^{\dagger}(\vec{r}+\hat{n})
U_{GH}[\vec{r},\hat{n}]\Psi_{H}(\vec{r})
\Psi_{I}^{\dagger}(\vec{r}+\hat{n})
U_{IL}[\vec{r},\hat{n}]\Psi_{L}(\vec{r})\right.\cr
&&\left.\sum_{\vec{r}}S(\vec{r})
\Psi_{M}^{\dagger}(\vec{r}+\hat{n})U_{MN}[\vec{r},\hat{n}]
\Psi_{N}(\vec{r})|0>\right\}\cr
&&-\frac{8a^2(N_C-1)}{g^6C_2^3(N_C-3)(N_C-2)}\left\{
<M^{\dagger}|H_q^{\dagger}H_q^{\dagger}H_q
\sum_{[\vec{r},\hat{n}]}\eta(\hat{n})\Psi_{I}^{\dagger}(\vec{r}+
\hat{n})U_{IL}[\vec{r},\hat{n}]\Psi_{L}(\vec{r})\right.\cr
&&\left.\sum_{\vec{r}}S(\vec{r})
\Psi_{M}^{\dagger}(\vec{r}+\hat{n})U_{MN}[\vec{r},\hat{n}]
\Psi_{N}(\vec{r})|0>\right.\cr
&&\left.+\frac{1}{a(N_C-4)}<M^{\dagger}|H_q^{\dagger}
H_q^{\dagger}\sum_{[\vec{r},\hat{n}]}\Psi_{G}^{\dagger}(\vec{r}
+\hat{n})U_{GH}[\vec{r},\hat{n}]\Psi_{G}(\vec{r})
\Psi_{I}^{\dagger}(\vec{r}+\hat{n})
U_{IL}[\vec{r},\hat{n}]\Psi_{L}(\vec{r})\right.\cr
&&\left.\sum_{\vec{r}}S(\vec{r})\Psi_{M}^{\dagger}(\vec{r}+\hat{n})
U_{MN}[\vec{r},\hat{n}]\Psi_{N}(\vec{r})|0>\right\}\cr
&&-\frac{4a(N_C-1)}{g^6C_2^3(N_C-3)^2(N_C-2)}\left\{
<0|\sum_{\vec{r}}S^{\star}(\vec{r})\Psi_{A}^{\dagger}(\vec{r})
U_{AB}^{\dagger}[\vec{r},\hat{n}]\Psi_{B}(\vec{r}+\hat{n})\right.\cr
&&\left.\sum_{[\vec{r},\hat{n}]}\eta(\hat{n})
\Psi_{C}^{\dagger}(\vec{r})U_{CD}^{\dagger}[\vec{r},\hat{n}]
\Psi_{D}(\vec{r}+\hat{n})H_q^{\dagger}
H_q\sum_{[\vec{r^{\prime}},\hat{m}]}\eta(\hat{m})
\Psi_{I}^{\dagger}(\vec{r^{\prime}}+\hat{m})
U_{IL}[\vec{r^{\prime}},\hat{m}]\Psi_{L}(\vec{r^{\prime}})\right.\cr
&&\left.\sum_{\vec{r^{\prime}}}S(\vec{r^{\prime}})
\Psi_{M}^{\dagger}(\vec{r^{\prime}}+\hat{m})
U_{MN}[\vec{r^{\prime}},\hat{m}]\Psi_{N}(\vec{r^{\prime}})|0>\right.\cr
&&\left.+\frac{1}{a(N_C-4)}
<0|\sum_{\vec{r}}S^{\star}(\vec{r})
\Psi_{A}^{\dagger}(\vec{r})U_{AB}^{\dagger}[\vec{r},\hat{n}]
\Psi_{B}(\vec{r}+\hat{n})\right.\cr
&&\left.\sum_{[\vec{r},\hat{n}]}\eta(\hat{n})
\Psi_{C}^{\dagger}(\vec{r})U_{CD}^{\dagger}[\vec{r},\hat{n}]
\Psi_{D}(\vec{r}+\hat{n})H_q^{\dagger}
\sum_{[\vec{r^{\prime}},\hat{m}]}\eta(\hat{m})
\Psi_{G}^{\dagger}(\vec{r^{\prime}}+\hat{m})
U_{GH}[\vec{r^{\prime}},\hat{m}]\Psi_{H}(\vec{r^{\prime}})\right.\cr
&&\left.\sum_{[\vec{r^{\prime}},\hat{m}]}\eta(\hat{m})
\Psi_{I}^{\dagger}(\vec{r^{\prime}}+\hat{m})
U_{IL}[\vec{r^{\prime}},\hat{m}]\Psi_{L}(\vec{r^{\prime}})
\sum_{\vec{r^{\prime}}}S(\vec{r^{\prime}})\Psi_{M}^{\dagger}
(\vec{r^{\prime}}+\hat{m})U_{MN}[\vec{r^{\prime}},\hat{m}]
\Psi_{N}(\vec{r^{\prime}})|0>\right\}.
\end{eqnarray}
\noindent The matrix elements in Eq.(\ref{UUU}) are evaluated by means
of the integrals over the gauge group elements given in Appendix C.

Group integration yields a nonvanishing result only if each link exhibits a
combination of matrices from which a color singlet may be
formed. Then, for each matrix element one should compute all the possible
integrals obtained by putting three $UU^{\dagger}$ pairs on a different
link [and then using Eq.(\ref{2U})] or two of them on the same link 
[Eqs.(\ref{2U}) and (\ref{4U})], 
or all the three on the same link [Eq.(\ref{6U})]. 
For $\rho$ this term gives 
\begin{eqnarray}
&&[(A)]_{\rho}=\cr
&&\frac{1}{4g^6aC_2^3}\left\{-86+\left(\frac{17}{2}N_C\right)N
-\frac{N_C^2}{4}N^2+\frac{4(N_C-1)(10N_C-49)}{(N_C-2)(N_C-4)}
-\frac{5N_C(N_C-1)}{2(N_C-2)}N\right.\cr
&&\left.+\frac{4}{N_C-3}
\left[-\frac{N_C(N_C-1)^2}{N_C-2}N
+\frac{4(N_C-1)^2(4N_C-19)}{(N_C-2)(N_C-4)}\right]\right.\cr
&&\left.+\frac{4(N_C-1)^2}{(N_C-3)^2}\left[-\frac{N_C}{2(N_C-2)}N
-\frac{2(N_C^2-8N_C+22)}{(N_C-2)(N_C-4)}\right]\right\}.
\end{eqnarray}
\noindent Now, one defines
\begin{equation}
\label{IV2B}
(B)=<M|H_q^{\dagger}\Lambda_{M}H_q^{\dagger}\Lambda_{M}
H_q\Lambda_{M}H_q|M^{\dagger}>.
\end{equation}
Combining Eq.(\ref{U2}) and the action of $\Pi_M$,
Eq.(\ref{IV2B}) becomes
\begin{eqnarray}
\label{IV2B1}
&&(B)=\frac{4a^2}{g^4C_2^2}\left[<M|H_q^{\dagger}H_q^{\dagger}
\Lambda_{M}H_qH_q|M^{\dagger}>\right.\cr
&&\left.+\frac{1}{16a^4}<M|\sum_{[\vec{r},\hat{n}]}
\Psi_{A}^{\dagger}(\vec{r})U_{AB}^{\dagger}[\vec{r},\hat{n}]
\Psi_{B}(\vec{r}+\hat{n})\Psi_{C}^{\dagger}(\vec{r})
U_{CD}^{\dagger}[\vec{r},\hat{n}]\Psi_{D}(\vec{r}+\hat{n})\right.\cr
&&\left.\Lambda_{M}\Psi_{E}^{\dagger}(\vec{r}+\hat{n})U_{EF}[\vec{r},\hat{n}]
\Psi_{F}(\vec{r})\Psi_{G}^{\dagger}(\vec{r}+\hat{n})U_{GH}[\vec{r},\hat{n}]
\Psi_{H}(\vec{r})|M^{\dagger}>\right].
\end{eqnarray}
\noindent Using again Eq.(\ref{F}), Eq.(\ref{F1}) and Eq.(\ref{M}),
one finds
\begin{equation}
\label{UUUc}
\Lambda_M(H_e)U_{AB}[\vec{r},\hat{n}]U_{CD}[\vec{r^{\prime}},\hat{m}]
U_{EF}^\dagger[\vec{r^{\prime\prime}},\hat{l}]|0>=-\frac{a}{g^2C_2}U
_{AB}[\vec{r},\hat{n}]
U_{CD}[\vec{r^{\prime}},\hat{m}]U_{EF}^\dagger[\vec{r^{\prime\prime}},\hat{l}]|0>
\end{equation}
\noindent and
\begin{eqnarray}
\label{UUUc1}
&&\Lambda_M(H_e)U_{AB}[\vec{r},\hat{n}]U_{CD}[\vec{r},\hat{n}]
U_{EF}^\dagger[\vec{r^{\prime}},\hat{l}]|0>=\cr
&&-\frac{2aN_C}{g^2(N_C+1)(N_C-2)}U_{AB}[\vec{r},\hat{n}]
U_{CD}[\vec{r},\hat{n}]U_{EF}^\dagger[\vec{r^{\prime}},\hat{l}]|0>.
\end{eqnarray}
\noindent Taking into account Eq.(\ref{UUUc}) and Eq.(\ref{UUUc1}),
one has
\begin{eqnarray}
\label{IV2B2}
&&(B)=-\frac{4a^3}{g^6C_2^3}\left[<M|H_q^{\dagger}H_q^{\dagger}
\Pi_{M}H_qH_q|M^{\dagger}>\right.\cr
&&\left.+\frac{(N_C-1)}{16a^4(N_C-2)}<M|\sum_{[\vec{r},\hat{n}]}
\Psi_{A}^{\dagger}(\vec{r})U_{AB}^{\dagger}[\vec{r},\hat{n}]
\Psi_{B}(\vec{r}+\hat{n})\Psi_{C}^{\dagger}(\vec{r})U_{CD}^{\dagger}
[\vec{r},\hat{n}]\Psi_{D}(\vec{r}+\hat{n})\right.\cr
&&\left.\Pi_{M}\Psi_{E}^{\dagger}(\vec{r}+\hat{n})U_{EF}[\vec{r},\hat{n}]
\Psi_{F}(\vec{r})\Psi_{G}^{\dagger}(\vec{r}+\hat{n})
U_{GH}[\vec{r},\hat{n}]\Psi_{H}(\vec{r})|M^{\dagger}>\right].
\end{eqnarray}
\noindent The two different combinations of link variables allowed for
the first term in Eq.(\ref{IV2B2}) are those where the gauge fields
of the meson operators $M$ and $M^{\dagger}$ are defined on the same
link. Then, the two possibilities may be represented by
\begin{eqnarray}
\nonumber
&&U_{AB}[\vec{r},\hat{n}]U_{CD}^{\dagger}[\vec{r^{\prime}},\hat{m}]
U_{EF}^{\dagger}[\vec{r^{\prime\prime}},\hat{l}]
U_{GH}[\vec{r^{\prime\prime}},\hat{l}]
U_{IL}[\vec{r^{\prime}},\hat{m}]U_{MN}^{\dagger}[\vec{r},\hat{n}]\cr
&&U_{AB}[\vec{r},\hat{n}]U_{CD}^{\dagger}[\vec{r^{\prime}},\hat{m}]
U_{EF}^{\dagger}[\vec{r^{\prime\prime}},\hat{l}]U_{GH}
[\vec{r^{\prime}},\hat{m}]U_{IL}[\vec{r^{\prime\prime}},\hat{l}]
U_{MN}^{\dagger}[\vec{r},\hat{n}]
\end{eqnarray}
where the link variables in this expression have the same ordering 
they have in Eq.(\ref{IV2B2}).
\noindent Since the gauge variables of the quark Hamiltonian are
on the same link, for the second term in Eq.(\ref{IV2B2}) there is
only one possibility, i.e., the gauge fields of the meson operators
must be defined on the same link,
$$U_{AB}[\vec{r},\hat{n}]U_{CD}^{\dagger}[\vec{r^{\prime}},\hat{m}]
U_{EF}^{\dagger}[\vec{r^{\prime}},\hat{m}]U_{GH}[\vec{r^{\prime}},\hat{m}]
U_{IL}[\vec{r^{\prime}},\hat{m}]U_{MN}^{\dagger}[\vec{r},\hat{n}].$$
After the integration over the gauge fields one gets
for $\rho$
\begin{equation}
[(B)]_{\rho}=\frac{1}{4g^6aC_2^3}\left[-72-12N_C+\frac{12(N_C-1)^2}{N_C-2}+
N\left(8N_C+\frac{N_C^2}{2}-\frac{N_C(N_C-1)^2}{2(N_C-2)}\right)-
\frac{N_C^2}{4}N^2\right].
\end{equation}
\noindent Analogously, the other terms of Eq.(\ref{7ter}) yield the
following contributions for $\rho$:
\begin{eqnarray}
\nonumber
&&<M|H_q\Lambda_{M}H_q^{\dagger}\Lambda_{M}H_q\Lambda_{M}
H_q^{\dagger}|M^{\dagger}>
=\frac{1}{4g^6aC_2^3}\left[-37-12N_C+N\left(4N_C+
\frac{N_C^2}{2}\right)\right.\cr
&&\left.-\frac{N_C^2}{8}N^2-\frac{(N_C-1)^3}{(N_C-2)(N_C-3)^2}\left(-12(N_C-1)
+\frac{N_C}{2}(N_C-1)N\right)\right]\cr
&&<M|H_q^{\dagger}\Lambda_{M}H_q\Lambda_{M}H_q^{\dagger}
\Lambda_{M}H_q|M^{\dagger}>
=\frac{1}{4g^6aC_2^3}\left[-40-12N_C+\frac{12(N_C-1)^2}{N_C-2}\right.\cr
&&\left.+N\left(4N_C+\frac{N_C^2}{2}
-\frac{N_C(N_C-1)^2}{2(N_C-2)}\right)-\frac{N_C^2}{8}N^2\right]\cr
&&<M|H_q\Lambda_{M}H_q^{\dagger}\Lambda_{M}H_q^{\dagger}
\Lambda_{M}H_q|M^{\dagger}>=\frac{1}{4g^6aC_2^3}\left[-34-12N_C
+\frac{12(N_C-1)^3}{(N_C-3)(N_C-2)}\right.\cr
&&\left.+N\left(4N_C+\frac{N_C^2}{2}
-\frac{N_C(N_C-1)^3}{2(N_C-3)(N_C-2)}\right)-\frac{N_C^2}{8}N^2\right]\cr
&&<M|H_q\Lambda_{M}H_q^{\dagger}\Lambda_{M}H_q^{\dagger}
\Lambda_{M}H_q^{\dagger}|M>=\frac{1}{4g^6aC_2^3}\left[16+4N_C
-\frac{4(N_C-1)^3}{(N_C-3)(N_C-2)}-N_CN\right]\cr
&&<M|H_q\Lambda_{M}H_q^{\dagger}\Lambda_{M}H_q^{\dagger}
\Lambda_{M}H_q^{\dagger}|M>
=\frac{1}{4g^6aC_2^3}\left[16+4N_C-\frac{4(N_C-1)^2}{N_C-2}-N_CN\right].
\end{eqnarray}
\noindent Using Eq.(\ref{U1}), Eq.(\ref{U2}), and the result
(\ref{so3}), the contribution of
Eq.(\ref{IV2}) to $E_{M}^{(4)}$ for $\rho$ may be easily computed
\begin{eqnarray}
&&\left(<M|H_q\frac{\Pi_{M}}{E_{M}^{(0)}-H_e}H_q|M><M|H_q
\frac{\Pi_{M}}{(E_{M}^{(0)}-H_e)^2}H_q|M>\right)_{\rho}\cr
&&=\frac{2}{g^6aC_2^3}\left[\frac{3N_C-13}{N_C-3}-\frac{N_C}{2}N\right]
\left[-\frac{N_C^2-18N_C+37}{(N_C-3)^2}+\frac{N_C}{2}N\right].
\end{eqnarray}
\noindent Collecting all these contributions, one finds that the
$N$-dependent terms cancel against the quark Hamiltonian contribution
to $E_{0}^{(4)}$ [see Eq.(\ref{T})] and the remaining terms are
$N$-independent. This is a very good check of our computation since the
meson mass should be an intensive quantity. 

Adding up all the $N$-independent terms
the contribution due to the quark Hamiltonian is given by
\begin{equation}
\label{IVr}
E_{\rho}^{(4)}-E_{q0}^{(4)}=\frac{31180-48674N_C+29051N_C^2-8183N_C^3
+1069N_C^4-51N_C^5}{2g^6aC_2^3(N_C-4)(N_C-3)^3(N_C-2)}
\end{equation}
where $E_{q0}^{(4)}$ is the contribution to the fourth order vacuum energy
due to the sole quark Hamiltonian (\ref{T}).
\noindent The results for the other mesons are listed below (see also
Appendix D  for the results for each matrix element):
\begin{eqnarray}
\label{IVp}
&&E_{\pi}^{(4)}-E_{q0}^{(4)}=\frac{29452-45650N_C+29963N_C^2-7471N_C^3
+949N_C^4-43N_C^5}{2g^6aC_2^3(N_C-4)(N_C-3)^3(N_C-2)}\\
\label{IVo}
&&E_{\omega}^{(4)}-E_{q0}^{(4)}=\frac{29452-45650N_C+26963N_C^2-7471N_C^3
+949N_C^4-43N_C^5}{2g^6aC_2^3(N_C-4)(N_C-3)^3(N_C-2)}\\
\label{IVb}
&&E_{b_1}^{(4)}-E_{q0}^{(4)}=\frac{36172-56386N_C+33803N_C^2-9671N_C^3
+1309N_C^4-67N_C^5}{2g^6aC_2^3(N_C-4)(N_C-3)^3(N_C-2)}\\
\label{IVa1}
&&E_{a_1}^{(4)}-E_{q0}^{(4)}=\frac{124972-205002N_C+130895N_C^2-40619N_C^3
+6121N_C^4-359N_C^5}{2g^6aC_2^3(N_C-4)(N_C-3)^3(N_C-2)}\\
\label{IVf2}
&&E_{f_2}^{(4)}-E_{q0}^{(4)}=\frac{29452-45650N_C+26963N_C^2-7471N_C^3
+949N_C^4-43N_C^5}{2g^6aC_2^3(N_C-4)(N_C-3)^3(N_C-2)}\\
\label{IVf0}
&&E_{f_0}^{(4)}-E_{q0}^{(4)}=\frac{76396-127034N_C+83111N_C^2-26803N_C^3
+4273N_C^4-271N_C^5}{2g^6aC_2^3(N_C-4)(N_C-3)^3(N_C-2)}
\end{eqnarray}
Equation (\ref{IV3}) yields the magnetic contribution to $E_{M}^{(4)}$ and it
is at the second order in the strong coupling expansion. This term
gives the same result for all the mesons and reads
\begin{equation}
\label{mag4}
(E_{M}^{(2)})_{magnetic}=-\frac{N}{g^6aC_2}+\frac{2}{g^6a(2N_C^2-1)(N_C^2-1)}
+\frac{2}{g^6a(2N_C-3)(N_C^2-1)}
\end{equation}
\noindent Again the $N$-dependent term in Eq.(\ref{mag4}) vanishes in
the difference between Eq.(\ref{mag4}) and the magnetic contribution
to the fourth order correction to the vacuum energy $E^{(4)}_{m0}$, 
Eq.(\ref{T}).
As for the second order, the fourth order correction to the
meson energy is finite and well defined after rescaling the coupling
constant $g^2N_C\rightarrow g^2$ and taking the 't Hooft limit (large
$N_C$ with $g^2N_C$ fixed).

\subsection{The meson spectrum}

 The lattice excitation masses are given by subtracting the energy of
 the ground state from the energies of the excitations
$$m_M=E_M-E_0$$
\noindent
Using the results of Eqs.(\ref{so3})-(\ref{so7}) and of
Eqs.(\ref{IVr})-(\ref{mag4}), one gets in the large $N_C$ limit
\begin{eqnarray}
\label{m1}
&&m_{\pi_0}=\frac{g^2}{a} (\frac{1}{4}+6 \epsilon^2 -171 \epsilon^4)\\
\label{m2}
&&m_{\rho}=\frac{g^2}{a} (\frac{1}{4}+6 \epsilon^2 -203 \epsilon^4)\\
\label{m3}
&&m_{\omega}=\frac{g^2}{a} (\frac{1}{4}+6 \epsilon^2 -171 \epsilon^4)\\
\label{m4}
&&m_{b_1}=\frac{g^2}{a}(\frac{1}{4}+10 \epsilon^2 -267 \epsilon^4)\\
\label{m5}
&&m_{a_1}=\frac{g^2}{a}(\frac{1}{4}+14 \epsilon^2 -1435 \epsilon^4)\\
\label{m6}
&&m_{f_2}=\frac{g^2}{a}(\frac{1}{4}+14 \epsilon^2 -875 \epsilon^4)\\
\label{m7}
&&m_{f_0}=\frac{g^2}{a}(\frac{1}{4}+18 \epsilon^2 -1083 \epsilon^4)
\end{eqnarray}
where $\epsilon=1/g^2$ and with $g$ we indicate the rescaled 
coupling constant $g^2N_C\to g^2$.
\noindent Equations(\ref{m1})-(\ref{m7}) provide the value of 
the meson masses up 
to the fourth order in the strong coupling expansion.

\section{Lattice vs. continuum}

The series given in the preceding section are derived for large $g^2$.
Since from renormalization group arguments for an
asymptotically free theory
$g^2=-c/\ln a$
for small $a$, the series for the meson masses
are valid only for large lattice spacings. 
To compare the results of the strong coupling
expansion with the continuum theory one needs some method of continuing
the series to the region in which $g^2=0$, i.e., $\epsilon=\infty$.
To make this extrapolation possible, it
is customary to make use of Pad\'e approximants \cite{Baker},
which allows one to
extrapolate a series expansion beyond the convergence radius. For this 
purpose one should consider the mass ratios, expand them as power series in
$y=\epsilon^2$ and then use $[1,1]$ Pad\'e approximants by writing the
mass ratios in the form
$$P^1_1=\frac{1+ay}{1+by}$$
\noindent where $a$ and $b$ are determined by expanding to order $y^2$
and equating coefficients. In the continuum limit this ratios yields
$a/b$. 

The results obtained with this method for the mass ratios of
the single link mesons are listed in Table 1.
\begin{table}[htbp]
\begin{center}
\caption{Mass ratios of the single-link mesons}
\vspace{.1in}
\begin{tabular}{lccc}
\hline Mass ratios & Our results & Banks et al. & Experimental values
\rule{0in} {4ex}\\[2ex] \hline $\frac{m_{\pi}}{m_{b_1}}$ & 0.75 & 0.86
& 0.11\rule{0in}{4ex}\\[2ex] $\frac{m_{\omega}}{m_{b_1}}$ & 0.75 &
0.86 & 0.63\rule{0in}{4ex}\\[2ex] $\frac{m_{\rho}}{m_{b_1}}$ & 0.71 &
0.81 & 0.62\rule{0in}{4ex}\\[2ex] 
$\frac{m_{b_1}}{m_{f_0}}$ & 0.82 &
1.03 & 0.90\rule{0in}{4ex}\\[2ex]
$\frac{m_{b_1}}{m_{a_1}}$ & 0.95 &
0.93 & 0.98\rule{0in}{4ex}\\[2ex] $\frac{m_{b_1}}{m_{f_2}}$ & 0.92 &
1.00 & 0.97\rule{0in}{4ex}\\[2ex]\hline
\end{tabular}
\end{center}
\end{table} 
For each mass ratio considered here, the $[1,1]$ Pad\'e approximant exists for
positive values for $a$ and $b$. Therefore, the extrapolation from $y=0$
to $y=\infty$ is singularity free in our approximation.

We considered the ratios between meson masses involving the $b_1$ meson.
For these ratios the results are in very good agreement with the
well known experimental values except for the pion mass.
This foreseeable failure is due to the lack of full chiral symmetry in 
the theory for large lattice
spacing. The $\pi$-$\rho$ splitting is so tiny because of the lack 
of significant spin-spin forces in the first four orders 
of strong-coupling perturbation theory. 
Magnetic field effects, loops of flux, are just not important through
this order. However, at sixth and eighth order such effects are important 
and should provide improved results. 

The mass of $f_0$ seems to be quite large but in agreement with the
experimental value of the meson $f_0(1370)$.

Our results show that the large $N_C$ expansion
provides a very good and systematic theoretical setting to evaluate physical
quantities of phenomenological interest.

\section{Concluding remarks}

In this paper we used the strong-coupling expansion of lattice QCD to compute 
-for large $N_C$- the low-lying unflavored meson spectrum. 
Our large $N_C$ Hamiltonian approach with staggered fermions evidences
that the possible ground states of strongly coupled lattice QCD are those of a 
spin $N_C/2$ antiferromagnetic Ising model. Choosing one of the two
ground states amounts then to the spontaneous breaking of
the discrete chiral symmetry
corresponding to translations by a lattice site.
As a consequence a nonvanishing chiral condensate should arise; 
it would then be 
interesting to compute -within our formalism- the chiral condensate 
on the lattice and then compare it with the results of numerical simulations.

Mesons are created by operators that, acting on the vacuum, create $q\bar q$ 
states with the desired quantum numbers. 
Their energy is computed up to the fourth order
in the strong coupling expansion. The meson masses are obtained by 
subtracting the vacuum state energy from the energies of the excitations.
After rescaling the coupling constant $g^2N_C\to g^2$ , according to 
the 't Hooft prescription, and taking the $N_C\to\infty$ limit
one finds that the series for the meson masses are well defined. 
Since it is expected that the continuum limit occurs without any phase transitions,
the series are analytically continued by using Pad\'e approximants. 

With the exception of the pion, which turns out to be degenerate in mass 
with $\omega$, the results we obtained are, already at the fourth order,
in good agreement with observed values.
Higher orders of the strong coupling expansion should evidence the 
splitting between the masses of $\pi_0$ and $\omega$ since the effects of the
magnetic field would be more pronounced.

The strong-coupling limit of QCD is not universal, thus adding
an irrelevant operator to the Hamiltonian leads to the same physical 
predictions in the continuum limit. This allows one to introduce
arbitrary parameters, the coefficients of these irrelevant operators,
which are then fixed by fitting the experimental data.
We have shown that, in the 
't Hooft limit, the meson masses can be 
computed without introducing irrelevant operators and consequently 
without arbitrary parameters.
Irrelevant operators are usually introduced~\cite{Banks:1976ia}, in fact,
mainly because in this way the meson masses are well defined for $N_C=3$.
Our large $N_C$ approach does not need any irrelevant operator;
it yields series expansions for the meson masses
which are free of divergences and thus well defined.

We determined the ratios between 
two single link mesons
even if these ratios are expected to be less reliable than the ratios involving
meson masses relative to the nucleon mass. The reason is that all the mesons
are degenerate at zeroth order so small differences of second and fourth 
Taylor series coefficients control the $[1,1]$ Pad\'e approximants, 
and a considerable amount of information in the series is lost. The good 
agreement between our results and the experimental values for the ratios
between meson masses shows that the large $N_C$ limit is very effective in
the strong-coupling region also. 
It should be interesting to extend our procedure
to provide a large $N_C$ evaluation of the ratios between the meson 
masses and the nucleon mass.

\begin{acknowledgments}
One of us, D.M., gratefully thanks Benjamin Svetitsky for helpful discussions.
\end{acknowledgments}

\appendix
\section{The staggered fermions formalism}

In this appendix the well known formalism for staggered
fermions originally developed by Kogut and Susskind~\cite{Kogut:ag} 
is reviewed. 
Other formulations exist in literature 
~\cite{Kluberg-Stern:1983dg} but in
this paper we use the notation and the conventions of
refs.~\cite{Banks:1976ia,Susskind:1976jm}. 

In this approach the two nonstrange
quark fields $u$ and $d$ are represented by a single-component 
lattice field $\Psi(\vec r)$. In a three-dimensional cubic lattice with
sites labeled by triplets of indices $\vec r=(x,y,z)$ one may 
define the lattice derivative operator
\be
\label{lattder}
\nabla_i\psi(\vec r)=\frac{1}{2a}\left[\psi(\vec r+\hat i)
-\psi(\vec r-\hat i)\right].
\ee
With the definition (\ref{lattder}) the massless Dirac equation
$$\dot \psi=-\vec \alpha \cdot \vec \nabla \psi$$
becomes, on the lattice,
\be
\label{Dirac}
\dot \psi(\vec r)=-\frac{1}{2a}\sum_{\vec n}\vec \alpha\cdot\hat n
[\psi(\vec r+\hat n)-\psi(\vec r-\hat n)]\ .
\ee
In Eq.(\ref{Dirac}) $\vec{\alpha}$ is the vector of Dirac matrices in the
representation
\begin{equation}
\label{alpha}
\gamma_{0}=\left(\begin{array}{cc} 1 & 0 \\ 0 & -1
\end{array}\right) \qquad 
\vec{\alpha}=\left(\begin{array}{cc} 0 & \vec{\sigma}\\ \vec{\sigma} & 0
\end{array}\right)
\end{equation} 
The dispersion law for Eq.(\ref{Dirac}) is given by
$$\omega=(\alpha_{x}\sin l_{x}+\alpha_{y}\sin l_{y}+\alpha_{z}\sin
l_{z})/a$$ and the low-energy spectrum is found at the eight corners
of a cube in $k$ space. 

It is customary to remove the degeneracy by reducing the
degrees of freedom. One may consider a one-component field, $\phi(r)$, per
lattice site which will represent a component of the Dirac filed. 
The lattice is subdivided into four sublattices to
place the four components of a conventional Dirac field and the corners of 
a unit cube are labeled as shown in Fig.1.
\begin{figure}
\includegraphics*{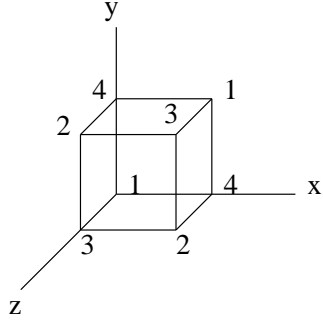}
\caption{\label{fig1}\small {\bf{Labeling of lattice sites}}}
\end{figure}
By further subdividing the lattice in sublattices there are
two complete and independent Dirac fields which exhaust the
low-frequency spectrum of the Dirac equation.
The lattice is now divided into a sublattice of sites for which $y$ is even
and another of sites for which $y$ is odd.
The fields are conveniently relabeled as follows
\begin{eqnarray}
\nonumber && \psi_i=f_i \quad (y=\rm{even})\\ \nonumber && \psi_i=g_i
\quad (y=\rm{odd}).
\end{eqnarray}
Writing Eq.(\ref{Dirac}) in Fourier transformed variables, one finds for $f$
and $g$ 
\begin{eqnarray}
\label{b5a}
&& \omega a\dot{f}=(\alpha_{z}\sin k_{z}+\alpha_{x}\sin k_{x})f+
(\alpha_{y}\sin k_{y})g\\
\label{b1}
&& \omega a\dot{g}=(\alpha_{z}\sin k_{z}+\alpha_{x}\sin k_{x})g+ 
(\alpha_{y}\sin k_{y})f.
\end{eqnarray}
There are two combinations of $f$ and $g$ which are
conventional Dirac fields for long wavelengths
\begin{equation}
\label{ufg}
u_i=f_i+g_i \quad (i=1,2,3,4)
\end{equation}
and
\begin{eqnarray}
\label{dfg}
&& d_1=f_2-g_2\cr 
&& d_2=-(f_1-g_1)\cr 
&& d_3=-(f_4-g_4)\cr
&& d_4=f_3-g_3
\end{eqnarray}
This is easily seen by summing up Eqs.(\ref{b5a}) and (\ref{b1}). 
One finds for the
component of the $u$ field in the representation (\ref{alpha})
\bea
\label{ueq}
&& a\omega \dot u_1(\vec k)=\sin k_x u_4-i \sin k_y u_4+\sin k_z u_3\cr
&& a\omega \dot u_2(\vec k)=\sin k_x u_3+i \sin k_y u_3-\sin k_z u_4\cr
&& a\omega \dot u_3(\vec k)=\sin k_x u_2-i \sin k_y u_2+\sin k_z u_1\cr
&& a\omega \dot u_4(\vec k)=\sin k_x u_1+i \sin k_y u_1-\sin k_z u_2
\eea
and identical equations for the $d$'s. If one now considers those normal 
modes with $\vec k$ small and set $\vec k=a\vec K$,
Eq.(\ref{ueq}), for small $\vec k$, takes the form
\bea
\label{smallk}
&& \omega \dot u_1(\vec K)=K_x u_4-i K_y u_4+K_z u_3\cr
&& \omega \dot u_2(\vec K)=K_x u_3+i K_y u_3-K_z u_4\cr
&& \omega \dot u_3(\vec K)=K_x u_2-i K_y u_2+K_z u_1\cr
&& \omega \dot u_4(\vec K)=K_x u_1+i K_y u_1-K_z u_2
\eea
and the same with $u \longleftrightarrow d$.
Equations(\ref{smallk}) are the normal Dirac equations for $u$ and $d$ 
in the representation (\ref{alpha}).

One may now turn to the free massless Dirac Hamiltonian in the continuum 
$$H=-i(u^{\dagger}\alpha_{i}\partial_{i}u+d^{\dagger}\alpha_{i}\partial_{i}d)$$
and write it in the lattice form by means of the definitions (\ref{ufg}) 
and (\ref{dfg}) for the $u$ and $d$ fields in the representation (\ref{alpha}).
In terms of the one component field $\phi(\vec r)$ one gets 
\begin{eqnarray}
\label{b2}
H=-\frac{i}{2a}\sum_{\vec{r}} &&
\left[(\phi^{\dagger}(\vec{r})\phi(\vec{r}+\hat{z})-h.c.)(-1)^{x+y}\right.\cr
&& \left.+(\phi^{\dagger}(\vec{r})\phi(\vec{r}+\hat{x})-h.c.)\right.\cr
&& \left.-i(\phi^{\dagger}(\vec{r})\phi(\vec{r}+\hat{y})+h.c.)(-1)^{x+y}\right].
\end{eqnarray}

Equation (\ref{b2}) may be written in a more symmetric form by defining
the following functions on the lattice
\begin{eqnarray}
\label{b6}
&& D(x,y)=\frac{1}{2}[(-1)^{x}+(-1)^{y}+(-1)^{x+y+1}+1]\\
\label{b7}
&& A(n)=\frac{1}{\sqrt{2}}[i^{n-1/2}+(-1)^{n-1/2}].
\end{eqnarray}
On the lattice sites $D$ and $A$ are either 1 or -1 and therefore
$$D^2=A^2=1.$$
Furthermore, $D$ and $A$ satisfy the relations
\bea
&& D(y,x)D(y,x+1)=(-1)^y\cr 
&& D(y+1,x+1)D(y,x)=(-1)^{x+y+1}\cr
&& A(y)A(y+1)=(-1)^y.
\eea
Defining the field
\begin{equation}
\label{b8}
\Psi(\vec{r})=(-1)^y(i)^{x+z}A(y)D(x,z)\phi(\vec{r})
\end{equation}
and using the previous definitions for $A$ and $D$ the lattice Hamiltonian becomes
\begin{eqnarray}
H_q=\frac{1}{2}\sum_{\vec{r}} &&
\left[\Psi^{\dagger}(\vec{r}+\hat{x})\Psi(\vec{r})(-1)^{z}\right.\cr
&& \left.\Psi^{\dagger}(\vec{r}+\hat{y})\Psi(\vec{r})(-1)^{x}\right.\cr
&& \left.\Psi^{\dagger}(\vec{r}+\hat{z})\Psi(\vec{r})(-1)^{y}+h.c.\right],
\end{eqnarray}
which is the quark Hamiltonian used in our paper.

\section{Construction of an eigenstate of $H_e$}

In this Appendix we explicit the construction of eigenstates of the
unperturbed Hamiltonian $H_e$ in order to evaluate the function of
$H_e$ in the perturbative expansion. Using Eq.(\ref{F}),
Eq.(\ref{F1}) and Eq.(\ref{M}), one finds
\begin{eqnarray}
\label{a1}
[H_e,U_{AB}^{\dagger}[\vec{r},\hat{n}]U_{CD}^{\dagger}
[\vec{r^{\prime}},\hat{m}]]|0>
&=&\frac{g^2}{a}\left(C_2U_{AB}^{\dagger}[\vec{r},\hat{n}]U_{CD}^{\dagger}
[\vec{r^{\prime}},\hat{m}]\right.\cr
&-&\left.\frac{N_C+1}{2N_C}U_{AB}^{\dagger}[\vec{r},\hat{n}]
U_{CD}^{\dagger}[\vec{r},\hat{n}]\delta([\vec{r},\hat{n}]
-[\vec{r^{\prime}},\hat{m}])\right)|0>.
\end{eqnarray}
Using Eq.(\ref{a1}), one finds
\begin{equation}
\label{a2}
[H_e,U_{AB}^{\dagger}[\vec{r},\hat{n}]U_{CD}^{\dagger}[\vec{r},\hat{n}]]|0>
=g^2\frac{(N_C-1)(N_C-2)}{2aN_C}U_{AB}^{\dagger}[\vec{r},\hat{n}]
U_{CD}^{\dagger}[\vec{r},\hat{n}]|0>.
\end{equation}

One may now look for an eigenstate of $H_e$ with eigenvalue
$g^2C_2/a$
\begin{eqnarray}
\label{a3}
&&H_e\left(U_{AB}^{\dagger}[\vec{r},\hat{n}]U_{CD}^{\dagger}
[\vec{r^{\dagger}},\hat{m}]+aU_{AB}^{\dagger}[\vec{r},\hat{n}]
U_{CD}^{\dagger}[\vec{r},\hat{n}]\right)|0>=\cr
&&\frac{g^2}{a}C_2\left(U_{AB}^{\dagger}[\vec{r},\hat{n}]U_{CD}^{\dagger}
[\vec{r^{\dagger}},\hat{m}]+aU_{AB}^{\dagger}[\vec{r},\hat{n}]
U_{CD}^{\dagger}[\vec{r},\hat{n}]\right)|0>
\end{eqnarray}
\noindent Taking into account Eq.(\ref{a1}) and Eq.(\ref{a2}) one gets
$a=-1$ and then the pertinent eigenstate is
\begin{eqnarray}
\label{a4}
&&H_e\left(U_{AB}^{\dagger}[\vec{r},\hat{n}]U_{CD}^{\dagger}
[\vec{r^{\dagger}},\hat{m}]-U_{AB}^{\dagger}[\vec{r},\hat{n}]
U_{CD}^{\dagger}[\vec{r},\hat{n}]\right)|0>=\cr
&&\frac{g^2}{a}C_2\left(U_{AB}^{\dagger}[\vec{r},\hat{n}]U_{CD}^{\dagger}
[\vec{r^{\dagger}},\hat{m}]-U_{AB}^{\dagger}[\vec{r},\hat{n}]
U_{CD}^{\dagger}[\vec{r},\hat{n}]\right)|0>
\end{eqnarray}
\noindent Using Eq.(\ref{a4}), one may evaluate the function
of $H_e$ appearing in the perturbative expansion
$$f(H_e)=\frac{\Pi_0}{E_{0}^{(0)}-H_e}.$$
\noindent From Eq.(\ref{a2}), one has
\begin{equation}
\label{a5}
f(H_e)U_{AB}^{\dagger}[\vec{r},\hat{n}]U_{CD}^{\dagger}[\vec{r},\hat{n}]|0>
=-\frac{2aN_C}{g^2(N_C+1)(N_C-2)}U_{AB}^{\dagger}[\vec{r},\hat{n}]
U_{CD}^{\dagger}[\vec{r},\hat{n}]|0>.
\end{equation}
\noindent Using Eq.(\ref{a5}), one gets from Eq.(\ref{a4})
\begin{eqnarray}
\label{a6}
&&f(H_e)U_{AB}^{\dagger}[\vec{r},\hat{n}]U_{CD}^{\dagger}
[\vec{r^{\prime}},\hat{m}]|0>=-\frac{a}{g^2C_2}\left(U_{AB}^{\dagger}
[\vec{r},\hat{n}]U_{CD}^{\dagger}[\vec{r^{\prime}},\hat{m}]\right.\cr
&&\left.+\frac{1}{N_C-2}U_{AB}^{\dagger}[\vec{r},\hat{n}]
U_{CD}^{\dagger}[\vec{r},\hat{n}]\delta([\vec{r},\hat{n}]
-[\vec{r^{\prime}},\hat{m}])\right)|0>.
\end{eqnarray}

\section{Integration over $SU(N_C)$}

In this appendix a table of the integrals over the group
elements of $SU(N_C)$ needed in the paper is provided. 

It is well known
that a basic ingredient to formulate $QCD$ on a lattice is to define
the measure of integration over the gauge degrees of freedom. Unlike
the continuum gauge fields, the lattice gauge fields are $SU(N_C)$
matrices with elements bounded in the range $[0,1]$; Wilson
\cite{Wilson:1974sk} proposed an invariant group measure, the Haar
measure, for the integration over the group elements. The integral is
defined so that, for any elements $g_1$ and $g_2$ of the group, one has
\begin{equation}
\label{APP1}
\int dU f(U)=\int dU f(Ug_1)=\int dU f(g_2U),
\end{equation}
\noindent with $f(U)$ a generic function over the group. When used in
nonperturbative studies of gauge theory, the definition (\ref{APP1})
avoids the problem of introducing a gauge fixing, since the field
variables are compact. The measure is normalized as
\begin{equation}
\int dU=1.
\end{equation}
\noindent The strong coupling expansion for an $SU(N_C)$ gauge theory
depends on the following identities for integration over link matrices
\cite{Creutz:ub}
\begin{eqnarray}
\label{1U}
&&\int dU U_{ab}=\int dU U_{ab}^{\dagger}=0\\\label{2U} &&\int dU
U_{ab}U_{cd}^{\dagger}=\frac{1}{N_C}\delta_{ad}\delta_{bc}\\
\nonumber
&&\int dU
U_{ab}U_{cd}^{\dagger}U_{ef}U_{gh}^{\dagger}=\frac{1}{N_C^2-1}
(\delta_{ad}\delta_{bc}
\delta_{eh}\delta_{fg}+\delta_{ah}\delta_{bg}\delta_{cf}\delta_{de})\\\label{4U}
&&\phantom{\int dU U_{ab}U_{cd}^{\dagger}U_{ef}U_{gh}^{\dagger}}
-\frac{1}{N_C(N_C^2-1)}(\delta_{ad}\delta_{bg}\delta_{eh}\delta_{fc}+\delta_{ah}\delta_{bc}\delta_{ed}\delta_{fg}).
\end{eqnarray}
\noindent One also needs the group integral over six elements, which
occurs at the fourth order in the strong coupling expansion of the
mass spectrum. The pertinent integral is \cite{Berruto:1999ir}
\begin{eqnarray}
\label{6U}
&&\int dU
U_{ab}U_{cd}^{\dagger}U_{ef}U_{gh}^{\dagger}U_{il}U_{mn}^{\dagger}=
\frac{N_C^2-2}{N_C(N_C^2-1)(N_C^2-4)}(\delta_{ad}\delta_{eh}
\delta_{in}\delta_{bc}\delta_{fg}\delta_{lm}
+\delta_{ad}\delta_{en}\delta_{hi}\delta_{bc}\delta_{fm}\delta_{gl}\cr
&&+\delta_{ah}\delta_{de}\delta_{in}\delta_{bg}\delta_{cf}\delta_{lm}
+\delta_{an}\delta_{di}\delta_{eh}\delta_{bm}\delta_{cl}\delta_{fg}
+\delta_{an}\delta_{de}\delta_{hi}\delta_{bm}\delta_{cf}\delta_{gl}
+\delta_{ah}\delta_{di}\delta_{en}\delta_{bg}\delta_{cl}\delta_{fm})\cr
&&-\frac{1}{(N_C^2-1)(N_C^2-4)}(\delta_{ad}\delta_{en}\delta_{hi}
\delta_{bc}\delta_{fg}\delta_{lm}+\delta_{ah}\delta_{de}\delta_{in}
\delta_{bc}\delta_{fg}\delta_{lm}+\delta_{an}\delta_{di}\delta_{eh}
\delta_{bc}\delta_{fg}\delta_{lm}\cr
&&+\delta_{ad}\delta_{eh}\delta_{in}\delta_{bc}\delta_{fm}\delta_{gl}
+\delta_{an}\delta_{de}\delta_{hi}\delta_{bc}\delta_{fm}\delta_{gl}+
\delta_{ah}\delta_{di}\delta_{en}\delta_{bc}\delta_{fm}\delta_{gl}
+\delta_{ad}\delta_{eh}\delta_{in}\delta_{bg}\delta_{cf}\delta_{lm}\cr
&&+\delta_{an}\delta_{de}\delta_{hi}\delta_{bg}\delta_{cf}\delta_{lm}
+\delta_{ah}\delta_{di}\delta_{en}\delta_{bg}\delta_{cf}\delta_{lm}
+\delta_{ad}\delta_{eh}\delta_{in}\delta_{bm}\delta_{cl}\delta_{fg}
+\delta_{an}\delta_{de}\delta_{hi}\delta_{bm}\delta_{cl}\delta_{fg}\cr
&&+\delta_{ah}\delta_{di}\delta_{en}\delta_{bm}\delta_{cl}\delta_{fg}
+\delta_{ad}\delta_{en}\delta_{hi}\delta_{bm}\delta_{cf}\delta_{gl}+
\delta_{ah}\delta_{de}\delta_{in}\delta_{bm}\delta_{cf}\delta_{gl}
+\delta_{an}\delta_{di}\delta_{eh}\delta_{bm}\delta_{cf}\delta_{gl}\cr
&&+\delta_{ad}\delta_{en}\delta_{hi}\delta_{bg}\delta_{cl}\delta_{fm}
+\delta_{ah}\delta_{de}\delta_{in}\delta_{bg}\delta_{cl}\delta_{fm}
+\delta_{an}\delta_{di}\delta_{eh}\delta_{bg}\delta_{cl}\delta_{fm})\cr
&&+\frac{2}{N_C(N_C^2-1)(N_C^2-4)}(\delta_{an}\delta_{de}\delta_{hi}
\delta_{bc}\delta_{fg}\delta_{lm}+\delta_{ah}\delta_{di}\delta_{en}
\delta_{bc}\delta_{fg}\delta_{lm}
+\delta_{ah}\delta_{de}\delta_{in}\delta_{bc}\delta_{fm}\delta_{gl}\cr
&&+\delta_{an}\delta_{di}\delta_{eh}\delta_{bc}\delta_{fm}\delta_{gl}
+\delta_{ad}\delta_{en}\delta_{hi}\delta_{bg}\delta_{cf}\delta_{lm}
+\delta_{an}\delta_{di}\delta_{eh}\delta_{bg}\delta_{cf}\delta_{lm}
+\delta_{ad}\delta_{en}\delta_{hi}\delta_{bm}\delta_{cl}\delta_{fg}\cr
&&+\delta_{ah}\delta_{de}\delta_{in}\delta_{bm}\delta_{cl}\delta_{fg}
+\delta_{ad}\delta_{eh}\delta_{in}\delta_{bm}\delta_{cf}\delta_{gl}
+\delta_{ah}\delta_{di}\delta_{en}\delta_{bm}\delta_{cf}\delta_{gl}
+\delta_{ad}\delta_{eh}\delta_{in}\delta_{bg}\delta_{cl}\delta_{fm}\cr
&&+\delta_{an}\delta_{de}\delta_{hi}\delta_{bg}\delta_{cl}\delta_{fm}).
\end{eqnarray}

\section{Matrix elements}

In this appendix the matrix elements useful for the
computation of the $\pi$, $\omega$, $b_1$, $a_1$,$f_2$ and $f_0$
energies at the fourth order in the strong-coupling expansion are reported.
\begin{enumerate}
\item{$\pi$}
\bea
&&<M^{\dagger}|H_q^{\dagger}\Lambda_{M}H_q^{\dagger}
\Lambda_{M}H_q\Lambda_{M}H_q|M>=\cr
&&\frac{1}{4g^6aC_2^3}\left\{-76+\left(\frac{17}{2}N_C\right)N
-\frac{N_C^2}{4}N^2-\frac{(N_C-1)}{N_C-2}
\left[\frac{5}{2}N_CN-58\right]\right.\cr
&&\left.-\frac{18(N_C-1)}{N_C-4}-\frac{4(N_C-1)^2}{N_C-3}
\left[\frac{(N_CN-22)}{N_C-2}+\frac{6}{N_C-4}\right]\right.\cr
&&\left.-\frac{4(N_C-1)^2}{(N_C-3)^2}
\left[\frac{N_C/2N+2N_C-14}{N_C-2}+\frac{6}{N_C-4}\right]\right\}
\eea
\bea
&&<M|H_q^{\dagger}\Lambda_{M}H_q^{\dagger}\Lambda_{M}H_q\Lambda_{M}
H_q|M^{\dagger}>=\frac{1}{4g^6aC_2^3}\left[-72-12N_C\right.\cr
&&\left.+\frac{12(N_C-1)^2}{N_C-2}
+N\left(8N_C+\frac{N_C^2}{2}-
\frac{N_C(N_C-1)^2}{2(N_C-2)}\right)-\frac{N_C^2}{4}N^2\right]
\eea
\bea
&&<M|H_q\Lambda_{M}H_q^{\dagger}\Lambda_{M}H_q\Lambda_{M}
H_q^{\dagger}|M^{\dagger}>=\frac{1}{4g^6aC_2^3}\left[-33-12N_C
-\frac{N_C^2}{8}N^2\right.\cr
&&\left.+N\left(4N_C+\frac{N_C^2}{2}\right)
-\frac{(N_C-1)^4}{(N_C-2)(N_C-3)^2}\left(-12+\frac{N_C}{2}N\right)\right]
\eea
\bea
&&<M|H_q^{\dagger}\Lambda_{M}H_q\Lambda_{M}H_q^{\dagger}
\Lambda_{M}H_q|M^{\dagger}>=\frac{1}{4g^6aC_2^3}\left[-38-12N_C\right.\cr
&&\left.+\frac{12(N_C-1)^2}{N_C-2}
+N\left(4N_C+\frac{N_C^2}{2}
-\frac{N_C(N_C-1)^2}{2(N_C-2)}\right)-\frac{N_C^2}{8}N^2\right]
\eea
\bea
&&<M|H_q\Lambda_{M}H_q^{\dagger}\Lambda_{M}H_q^{\dagger}
\Lambda_{M}H_q|M^{\dagger}>=\frac{1}{4g^6aC_2^3}\left[-32-12N_C
+\frac{12(N_C-1)^3}{(N_C-3)(N_C-2)}\right.\cr
&&\left.+N\left(4N_C+\frac{N_C^2}{2}
-\frac{N_C(N_C-1)^3}{2(N_C-3)(N_C-2)}\right)-\frac{N_C^2}{8}N^2\right]
\eea
\be
<M|H_q\Lambda_{M}H_q^{\dagger}\Lambda_{M}H_q^{\dagger}\Lambda_{M}
H_q^{\dagger}|M>=
\frac{1}{4g^6aC_2^3}\left[14+4N_C-N_CN
-\frac{4(N_C-1)^3}{(N_C-3)(N_C-2)}\right]
\ee
\be
<M|H_q\Lambda_{M}H_q^{\dagger}\Lambda_{M}H_q^{\dagger}
\Lambda_{M}H_q^{\dagger}|M>=\frac{1}{4g^6aC_2^3}\left[12+4N_C
-\frac{4(N_C-1)^2}{N_C-2}-N_CN\right]
\ee
\bea
&&\left(<M|H_q\frac{\Pi_{M}}{E_{M}^{(0)}-H_e}H_q|M>
<M|H_q\frac{\Pi_{M}}{(E_{M}^{(0)}-H_e)^2}H_q|M>\right)_{\pi}\cr
&&=\frac{2}{g^6C_2^3}\left[\frac{3N_C-13}{N_C-3}-\frac{N_C}{2}N\right]
\left[-\frac{N_C^2-18N_C+37}{(N_C-3)^2}+\frac{N_C}{2}N\right]
\eea

\item{$\omega$}
\bea
&&<M^{\dagger}|H_q^{\dagger}\Lambda_{M}H_q^{\dagger}
\Lambda_{M}H_q\Lambda_{M}H_q|M>=\cr
&&\frac{1}{4g^6aC_2^3}\left\{-96
+\left(\frac{17}{2}N_C\right)N-\frac{N_C^2}{4}N^2
-\frac{(N_C-1)}{N_C-2}
\left[\frac{5}{2}N_CN-58\right]\right.\cr
&&\left.-\frac{18(N_C-1)}{N_C-4}-\frac{4(N_C-1)^2}{N_C-3}
\left[\frac{(N_CN-22)}{N_C-2}+\frac{6}{N_C-4}\right]\right.\cr
&&\left.-\frac{4(N_C-1)^2}{(N_C-3)^2}
\left[\frac{N_C/2N+2N_C-14}{N_C-2}+\frac{6}{N_C-4}\right]\right\}
\eea
\bea
&&<M|H_q^{\dagger}\Lambda_{M}H_q^{\dagger}\Lambda_{M}H_q
\Lambda_{M}H_q|M^{\dagger}>=\frac{1}{4g^6aC_2^3}\left[-72-12N_C
+\frac{12(N_C-1)^2}{N_C-2}\right.\cr
&&\left.+N\left(8N_C+\frac{N_C^2}{2}
-\frac{N_C(N_C-1)^2}{2(N_C-2)}\right)-\frac{N_C^2}{4}N^2\right]
\eea
\bea
&&<M|H_q\Lambda_{M}H_q^{\dagger}\Lambda_{M}H_q\Lambda_{M}H_q^{\dagger}
|M^{\dagger}>=\frac{1}{4g^6aC_2^3}\left[-41-12N_C+N\left(4N_C
+\frac{N_C^2}{2}\right)\right.\cr
&&\left.-\frac{N_C^2}{8}N^2
-\frac{(N_C-1)^4}{(N_C-2)(N_C-3)^2}\left(-12+\frac{N_C}{2}N\right)\right]
\eea
\bea
&&<M|H_q^{\dagger}\Lambda_{M}H_q\Lambda_{M}H_q^{\dagger}
\Lambda_{M}H_q|M^{\dagger}>=\frac{1}{4g^6aC_2^3}
\left[-42-12N_C+\frac{12(N_C-1)^2}{N_C-2}\right.\cr
&&\left.+N
\left(4N_C+\frac{N_C^2}{2}-\frac{N_C(N_C-1)^2}{2(N_C-2)}\right)
-\frac{N_C^2}{8}N^2\right]
\eea
\bea
&&<M|H_q\Lambda_{M}H_q^{\dagger}\Lambda_{M}H_q^{\dagger}
\Lambda_{M}H_q|M^{\dagger}>=\frac{1}{4g^6aC_2^3}
\left[-36-12N_C+\frac{12(N_C-1)^3}{(N_C-3)(N_C-2)}\right.\cr
&&\left.+N\left(4N_C+\frac{N_C^2}{2}
-\frac{N_C(N_C-1)^3}{2(N_C-3)(N_C-2)}\right)-\frac{N_C^2}{8}N^2\right]
\eea
\be
<M|H_q\Lambda_{M}H_q^{\dagger}\Lambda_{M}H_q^{\dagger}
\Lambda_{M}H_q^{\dagger}|M>=\frac{1}{4g^6aC_2^3}\left[22+4N_C-N_CN
-\frac{4(N_C-1)^3}{(N_C-3)(N_C-2)}\right]
\ee
\be
<M|H_q\Lambda_{M}H_q^{\dagger}\Lambda_{M}H_q^{\dagger}
\Lambda_{M}H_q^{\dagger}|M>=\frac{1}{4g^6aC_2^3}\left[24+4N_C
-\frac{4(N_C-1)^2}{N_C-2}-N_CN\right]
\ee
\bea
&&\left(<M|H_q\frac{\Pi_{M}}{E_{M}^{(0)}-H_e}H_q|M><M|H_q
\frac{\Pi_{M}}{(E_{M}^{(0)}-H_e)^2}H_q|M>\right)_{\omega}\cr
&&=\frac{2}{g^6aC_2^3}\left[\frac{3N_C-13}{N_C-3}-\frac{N_C}{2}
N\right]\left[-\frac{N_C^2-18N_C+37}{(N_C-3)^2}+\frac{N_C}{2}N\right]
\eea

\item{$b_1$}
\bea
&&<M^{\dagger}|H_q^{\dagger}\Lambda_{M}H_q^{\dagger}\Lambda_{M}
H_q\Lambda_{M}H_q|M>=\cr
&&\frac{1}{4g^6aC_2^3}\left\{-76+\left(\frac{17}{2}N_C\right)
N-\frac{N_C^2}{4}N^2
-\frac{(N_C-1)}{N_C-2}\left[\frac{5}{2}N_CN-58\right]\right.\cr
&&\left.-\frac{18(N_C-1)}{N_C-4}-\frac{4(N_C-1)^2}{N_C-3}\left[
\frac{(N_CN-22)}{N_C-2}+\frac{6}{N_C-4}\right]\right.\cr
&&\left.-\frac{4(N_C-1)^2}{(N_C-3)^2}
\left[\frac{N_C/2N+2N_C-14}{N_C-2}+\frac{6}{N_C-4}\right]\right\}
\eea
\bea
&&<M|H_q^{\dagger}\Lambda_{M}H_q^{\dagger}\Lambda_{M}H_q
\Lambda_{M}H_q|M^{\dagger}>=\frac{1}{4g^6aC_2^3}\left[-72-12N_C
+\frac{12(N_C-1)^2}{N_C-2}\right.\cr
&&\left.+N\left(8N_C+\frac{N_C^2}{2}
-\frac{N_C(N_C-1)^2}{2(N_C-2)}\right)-\frac{N_C^2}{4}N^2\right]
\eea
\bea
&&<M|H_q\Lambda_{M}H_q^{\dagger}\Lambda_{M}H_q\Lambda_{M}H_q^{\dagger}
|M^{\dagger}>=\frac{1}{4g^6aC_2^3}\left[-33-12N_C+N\left(4N_C
+\frac{N_C^2}{2}\right)\right.\cr
&&\left.-\frac{N_C^2}{8}N^2
-\frac{(N_C-1)^4}{(N_C-2)(N_C-3)^2}\left(-12+\frac{N_C}{2}N\right)\right]
\eea
\bea
&&<M|H_q^{\dagger}\Lambda_{M}H_q\Lambda_{M}H_q^{\dagger}
\Lambda_{M}H_q|M^{\dagger}>=\frac{1}{4g^6aC_2^3}\left[-38-12N_C
+\frac{12(N_C-1)^2}{N_C-2}\right.\cr
&&\left.+N\left(4N_C+\frac{N_C^2}{2}-\frac{N_C(N_C-1)^2}{2(N_C-2)}\right)
-\frac{N_C^2}{8}N^2\right]
\eea
\bea
&&<M|H_q\Lambda_{M}H_q^{\dagger}\Lambda_{M}H_q^{\dagger}
\Lambda_{M}H_q|M^{\dagger}>=\frac{1}{4g^6aC_2^3}\left[-32-12N_C
+\frac{12(N_C-1)^3}{(N_C-3)(N_C-2)}\right.\cr
&&\left.+N\left(4N_C+\frac{N_C^2}{2}
-\frac{N_C(N_C-1)^3}{2(N_C-3)(N_C-2)}\right)-\frac{N_C^2}{8}N^2\right]
\eea
\be
<M|H_q\Lambda_{M}H_q^{\dagger}\Lambda_{M}H_q^{\dagger}\Lambda_{M}
H_q^{\dagger}|M>=\frac{1}{4g^6aC_2^3}\left[-14-4N_C+N_CN
+\frac{4(N_C-1)^3}{(N_C-3)(N_C-2)}\right]
\ee
\be
<M|H_q\Lambda_{M}H_q^{\dagger}\Lambda_{M}H_q^{\dagger}
\Lambda_{M}H_q^{\dagger}|M>=\frac{1}{4g^6aC_2^3}\left[-12-4N_C
+\frac{4(N_C-1)^2}{N_C-2}+N_CN\right]
\ee
\bea
&&\left(<M|H_q\frac{\Pi_{M}}{E_{M}^{(0)}-H_e}H_q|M><M|H_q
\frac{\Pi_{M}}{(E_{M}^{(0)}-H_e)^2}H_q|M>\right)_{b_1}\cr
&&=\frac{2}{g^6aC_2^3}\left[\frac{5N_C-19}{N_C-3}-\frac{N_C}{2}N\right]
\left[-\frac{3N_C^2-30N_C+55}{(N_C-3)^2}+\frac{N_C}{2}N\right]
\eea

\item{$a_1$}
\bea
&&<M^{\dagger}|H_q^{\dagger}\Lambda_{M}H_q^{\dagger}\Lambda_{M}
H_q\Lambda_{M}H_q|M>=\cr
&&\frac{1}{4g^6aC_2^3}\left\{-91+\left(
\frac{17}{2}N_C\right)N-\frac{N_C^2}{4}N^2
-\frac{(N_C-1)}{N_C-2}\left[\frac{5}{2}N_CN-58\right]\right.\cr
&&\left.-\frac{18(N_C-1)}{N_C-4}-\frac{4(N_C-1)^2}{N_C-3}\left[
\frac{(N_CN-22)}{N_C-2}+\frac{6}{N_C-4}\right]\right.\cr
&&\left.-\frac{4(N_C-1)^2}{(N_C-3)^2}
\left[\frac{N_C/2N+2N_C-14}{N_C-2}+\frac{6}{N_C-4}\right]\right\}
\eea
\bea
&&<M|H_q^{\dagger}\Lambda_{M}H_q^{\dagger}\Lambda_{M}H_q\Lambda_{M}
H_q|M^{\dagger}>=\frac{1}{4g^6aC_2^3}\left[-72-12N_C
+\frac{12(N_C-1)^2}{N_C-2}\right.\cr
&&\left.+N\left(8N_C+\frac{N_C^2}{2}
-\frac{N_C(N_C-1)^2}{2(N_C-2)}\right)-\frac{N_C^2}{4}N^2\right]
\eea
\bea
&&<M|H_q\Lambda_{M}H_q^{\dagger}\Lambda_{M}H_q\Lambda_{M}H_q^{\dagger}
|M^{\dagger}>=\frac{1}{4g^6aC_2^3}\left[-39-12N_C+N\left(4N_C
+\frac{N_C^2}{2}\right)\right.\cr
&&\left.-\frac{N_C^2}{8}N^2
-\frac{(N_C-1)^4}{(N_C-2)(N_C-3)^2}\left(-12+\frac{N_C}{2}N\right)\right]
\eea
\bea
&&<M|H_q^{\dagger}\Lambda_{M}H_q\Lambda_{M}H_q^{\dagger}\Lambda_{M}
H_q|M^{\dagger}>=\frac{1}{4g^6aC_2^3}\left[-41-12N_C
+\frac{12(N_C-1)^2}{N_C-2}\right.\cr
&&\left.+N\left(4N_C+\frac{N_C^2}{2}
-\frac{N_C(N_C-1)^2}{2(N_C-2)}\right)-\frac{N_C^2}{8}N^2\right]
\eea
\bea
&&<M|H_q\Lambda_{M}H_q^{\dagger}\Lambda_{M}H_q^{\dagger}\Lambda_{M}
H_q|M^{\dagger}>=\frac{1}{4g^6aC_2^3}\left[-35-12N_C
+\frac{12(N_C-1)^3}{(N_C-3)(N_C-2)}\right.\cr
&&\left.+N\left(4N_C+\frac{N_C^2}{2}
-\frac{N_C(N_C-1)^3}{2(N_C-3)(N_C-2)}\right)-\frac{N_C^2}{8}N^2\right]
\eea
\be
<M|H_q\Lambda_{M}H_q^{\dagger}\Lambda_{M}H_q^{\dagger}\Lambda_{M}
H_q^{\dagger}|M>=\frac{1}{4g^6aC_2^3}[-3]
\ee
\be
<M|H_q\Lambda_{M}H_q^{\dagger}\Lambda_{M}H_q^{\dagger}\Lambda_{M}
H_q^{\dagger}|M>=\frac{1}{4g^6aC_2^3}[-4]
\ee
\bea
&&\left(<M|H_q\frac{\Pi_{M}}{E_{M}^{(0)}-H_e}H_q|M><M|H_q
\frac{\Pi_{M}}{(E_{M}^{(0)}-H_e)^2}H_q|M>\right)_{a_1}\cr
&&=\frac{2}{g^6aC_2^3}\left[\frac{7N_C-25}{N_C-3}-\frac{N_C}{2}N
\right]\left[-\frac{5N_C^2-42N_C+73}{(N_C-3)^2}+\frac{N_C}{2}N\right]
\eea

\item{$f_2$}
\bea
&&<M^{\dagger}|H_q^{\dagger}\Lambda_{M}H_q^{\dagger}\Lambda_{M}
H_q\Lambda_{M}H_q|M>=\cr
&&\frac{1}{4g^6aC_2^3}\left\{-86
+\left(\frac{17}{2}N_C\right)N-\frac{N_C^2}{4}N^2
-\frac{(N_C-1)}{N_C-2}\left[\frac{5}{2}N_CN-58\right]\right.\cr
&&\left.-\frac{18(N_C-1)}{N_C-4}-\frac{4(N_C-1)^2}{N_C-3}
\left[\frac{(N_CN-22)}{N_C-2}+\frac{6}{N_C-4}\right]\right.\cr
&&\left.-\frac{4(N_C-1)^2}{(N_C-3)^2}
\left[\frac{N_C/2N+2N_C-14}{N_C-2}+\frac{6}{N_C-4}\right]\right\}
\eea
\bea
&&<M|H_q^{\dagger}\Lambda_{M}H_q^{\dagger}\Lambda_{M}H_q\Lambda_{M}
H_q|M^{\dagger}>=\frac{1}{4g^6aC_2^3}\left[-72-12N_C
+\frac{12(N_C-1)^2}{N_C-2}\right.\cr
&&\left.+N\left(8N_C+\frac{N_C^2}{2}
-\frac{N_C(N_C-1)^2}{2(N_C-2)}\right)-\frac{N_C^2}{4}N^2\right]
\eea
\bea
&&<M|H_q\Lambda_{M}H_q^{\dagger}\Lambda_{M}H_q\Lambda_{M}
H_q^{\dagger}|M^{\dagger}>=\frac{1}{4g^6aC_2^3}\left[-37-12N_C
+N\left(4N_C+\frac{N_C^2}{2}\right)\right.\cr
&&\left.-\frac{N_C^2}{8}N^2
-\frac{(N_C-1)^4}{(N_C-2)(N_C-3)^2}\left(-12+\frac{N_C}{2}N\right)\right]
\eea
\bea
&&<M|H_q^{\dagger}\Lambda_{M}H_q\Lambda_{M}H_q^{\dagger}
\Lambda_{M}H_q|M^{\dagger}>=\frac{1}{4g^6aC_2^3}\left[-40-12N_C
+\frac{12(N_C-1)^2}{N_C-2}\right.\cr
&&\left.+N\left(4N_C+\frac{N_C^2}{2}
-\frac{N_C(N_C-1)^2}{2(N_C-2)}\right)-\frac{N_C^2}{8}N^2\right]
\eea
\bea
&&<M|H_q\Lambda_{M}H_q^{\dagger}\Lambda_{M}H_q^{\dagger}\Lambda_{M}
H_q|M^{\dagger}>=\frac{1}{4g^6aC_2^3}\left[-34-12N_C
+\frac{12(N_C-1)^3}{(N_C-3)(N_C-2)}\right.\cr
&&\left.+N\left(4N_C+\frac{N_C^2}{2}
-\frac{N_C(N_C-1)^3}{2(N_C-3)(N_C-2)}\right)-\frac{N_C^2}{8}N^2\right]
\eea
\be
<M|H_q\Lambda_{M}H_q^{\dagger}\Lambda_{M}H_q^{\dagger}\Lambda_{M}
H_q^{\dagger}|M>=\frac{1}{4g^6aC_2^3}\left[-18-4N_C+N_CN
+\frac{4(N_C-1)^3}{(N_C-3)(N_C-2)}\right]
\ee
\be
<M|H_q\Lambda_{M}H_q^{\dagger}\Lambda_{M}H_q^{\dagger}\Lambda_{M}
H_q^{\dagger}|M>=\frac{1}{4g^6aC_2^3}\left[-18-4N_C
+\frac{4(N_C-1)^2}{N_C-2}+N_CN\right]
\ee
\bea
&&\left(<M|H_q\frac{\Pi_{M}}{E_{M}^{(0)}-H_e}H_q|M>
<M|H_q\frac{\Pi_{M}}{(E_{M}^{(0)}-H_e)^2}H_q|M>\right)_{f_2}\cr
\nonumber
&&=\frac{2}{g^6aC_2^3}\left[\frac{7N_C-25}{N_C-3}
-\frac{N_C}{2}N\right]\left[-\frac{5N_C^2-42N_C+73}{(N_C-3)^2}
+\frac{N_C}{2}N\right]
\eea

\item{$f_0$}
\bea
&&<M^{\dagger}|H_q^{\dagger}\Lambda_{M}H_q^{\dagger}
\Lambda_{M}H_q\Lambda_{M}H_q|M>=\cr
&&\frac{1}{4g^6aC_2^3}
\left\{-181+\left(\frac{25}{2}N_C\right)N
-\frac{N_C^2}{4}N^2-\frac{(N_C-1)}{N_C-2}
\left[\frac{5}{2}N_CN-90\right]\right.\cr
&&\left.-\frac{18(N_C-1)}{N_C-4}-\frac{4(N_C-1)^2}{N_C-3}
\left[\frac{(N_CN-30)}{N_C-2}+\frac{6}{N_C-4}\right]\right.\cr
&&\left.-\frac{4(N_C-1)^2}{(N_C-3)^2}
\left[\frac{N_C/2N+2N_C-14}{N_C-2}+\frac{6}{N_C-4}\right]\right\}
\eea
\bea
&&<M|H_q^{\dagger}\Lambda_{M}H_q^{\dagger}\Lambda_{M}H_q
\Lambda_{M}H_q|M^{\dagger}>=\frac{1}{4g^6aC_2^3}\left[-72-12N_C
+\frac{12(N_C-1)^2}{N_C-2}\right.\cr
&&\left.+N\left(8N_C+\frac{N_C^2}{2}
-\frac{N_C(N_C-1)^2}{2(N_C-2)}\right)-\frac{N_C^2}{4}N^2\right]
\eea
\bea
&&<M|H_q\Lambda_{M}H_q^{\dagger}\Lambda_{M}H_q\Lambda_{M}
H_q^{\dagger}|M^{\dagger}>=\frac{1}{4g^6aC_2^3}\left[-70-12N_C
+N\left(5N_C+\frac{N_C^2}{2}\right)\right.\cr
&&\left.-\frac{N_C^2}{8}N^2
-\frac{(N_C-1)^4}{(N_C-2)(N_C-3)^2}\left(-12+\frac{N_C}{2}N\right)\right]
\eea
\bea
&&<M|H_q^{\dagger}\Lambda_{M}H_q\Lambda_{M}H_q^{\dagger}\Lambda_{M}
H_q|M^{\dagger}>=\frac{1}{4g^6aC_2^3}\left[-72-12N_C
+\frac{12(N_C-1)^2}{N_C-2}\right.\cr
&&\left.+N\left(5N_C+\frac{N_C^2}{2}-\frac{N_C(N_C-1)^2}{2(N_C-2)}\right)
-\frac{N_C^2}{8}N^2\right]
\eea
\bea
&&<M|H_q\Lambda_{M}H_q^{\dagger}\Lambda_{M}H_q^{\dagger}
\Lambda_{M}H_q|M^{\dagger}>=\frac{1}{4g^6aC_2^3}\left[-66-12N_C
+\frac{12(N_C-1)^3}{(N_C-3)(N_C-2)}\right.\cr
&&\left.+N\left(5N_C+\frac{N_C^2}{2}
-\frac{N_C(N_C-1)^3}{2(N_C-3)(N_C-2)}\right)-\frac{N_C^2}{8}N^2\right]
\eea
\be
<M|H_q\Lambda_{M}H_q^{\dagger}\Lambda_{M}H_q^{\dagger}
\Lambda_{M}H_q^{\dagger}|M>=\frac{1}{4g^6aC_2^3}\left[-66-6N_C+3N_CN
+\frac{6(N_C-1)^3}{(N_C-3)(N_C-2)}\right]
\ee
\be
<M|H_q\Lambda_{M}H_q^{\dagger}\Lambda_{M}H_q^{\dagger}\Lambda_{M}
H_q^{\dagger}|M>=\frac{1}{4g^6aC_2^3}\left[-72-6N_C
+\frac{6(N_C-1)^2}{N_C-2}+3N_CN\right]
\ee
\bea
&&\left(<M|H_q\frac{\Pi_{M}}{E_{M}^{(0)}-H_e}H_q|M>
<M|H_q\frac{\Pi_{M}}{(E_{M}^{(0)}-H_e)^2}H_q|M>\right)_{f_0}\cr
&&=\frac{2}{g^6aC_2^3}\left[\frac{9N_C-29}{N_C-3}
-\frac{N_C}{2}N\right]\left[-\frac{7N_C^2-54N_C+91}{(N_C-3)^2}
+\frac{N_C}{2}N\right]
\eea
\end{enumerate}

\end{document}